\documentclass[a4paper, 11pt]{article}
 
\usepackage[top=2.5cm, bottom=3.3cm, left=1.9cm, right=1.9cm]{geometry}

\usepackage{cite}
\usepackage{axodraw4j}
\usepackage{pstricks}
\usepackage{caption}
\usepackage{color}
\usepackage{multicol}
\usepackage{array}
\usepackage{amsmath,amsthm}
\usepackage{amsfonts}
\usepackage[latin1]{inputenc} 
\usepackage[T1]{fontenc}      
\usepackage{geometry}         
\usepackage[greek,english]{babel}  
\usepackage{graphicx}         
\usepackage{verbatim}         
\usepackage{slashed}
\usepackage{braket}

\title{\huge\bf
2D quantum gravity at three loops:\\ a \ct investigation
}
\author{L\ae titia \textsc{LEDUC} and Adel \textsc{BILAL} \\
\small \textit{Centre National de la Recherche Scientifique} \\
\small \textit{Laboratoire de Physique Théorique de l'\'Ecole Normale Supérieure} \\
\small \textit{24 rue Lhomond, F-75231 Paris Cedex 05, France} \\
\small \texttt{laetitia.leduc@lpt.ens.fr, adel.bilal@lpt.ens.fr}}
\date{}           

\numberwithin{equation}{section}

\newcommand\rfig[1]{Fig.\;\ref{#1}}
\newcommand\rtab[1]{Tab.\;\ref{#1}} 
\newcommand{\ct}{counterterm } 
\newcommand{\cts}{counterterms }
\newcommand{\im}[2]{\vcenter{\hbox{\includegraphics[scale=#1]{#2}}}}
\newcommand{\dx}{\mathrm{d}^2x\sqrt{g_*}}
\newcommand{\dnu}{\mathrm{d}\nu}
\newcommand{\dxy}[2]{\mathrm{d}^2#1\sqrt{g_*(#1)}\mathrm{d}^2#2\sqrt{g_*(#2)}}
\renewcommand{\textbf}[1]{\begingroup\bfseries\mathversion{bold}#1\endgroup}
\newcommand{\hk}{\widehat{\widetilde{K}}}
\newcommand{\kt}{\widetilde{K}}

\begin{document}
 
\maketitle

\noindent
We  analyse the divergences of the three-loop partition function at fixed area in 2D quantum gravity. Considering the Liouville action in the K\"ahler formalism, we extract the coefficient of the leading divergence $\sim A\Lambda^2 (\ln A\Lambda^2)^2$. This coefficient is non-vanishing. We discuss the counterterms one can and must add and compute their precise contribution to the partition function. This allows us to conclude that every local and non-local divergence in the partition function can be balanced by local counterterms, with the only exception of the maximally non-local divergence $(\ln A\Lambda^2)^3$. Yet, this latter is computed and does cancel  between the different three-loop diagrams. Thus, requiring locality of the counterterms is enough to renormalize the partition function. Finally, the structure of the new counterterms strongly suggests that they can be understood as a renormalization of the measure action.

\vskip.4cm
\hrule

\tableofcontents

\section{Introduction}
\label{sec:intro}

The coupling of conformal matter and two-dimensional quantum gravity is a subject which has been deeply studied, with a broad variety of approaches, from the discrete -- triangulations \cite{Ambjorn:1985az} and matrix models \cite{Douglas:1989ve} \cite{Gross:1989vs} \cite{Brezin:1990rb} \cite{Brezin:1989db} -- to the continuum approaches \cite{Knizhnik:1988ak} \cite{Distler:1988jt} \cite{David:1988hj}. In the continuum approach, most of the computations have been done within the conformal gauge. When the conformally coupled matter is integrated out, one ends up with the Liouville action as an effective gravity action. An interesting object to characterize is the partition function at fixed area $Z[A]$, where $A$ is the area of a Riemann surface of genus $h$. In the conformal gauge, $g=e^{2\sigma}g_0$, where $\sigma$ is the conformal factor and $g_0$ a background metric,  $Z[A]$ may be formally written as
\begin{align*}
Z[A]=\int \mathcal{D}\sigma~\exp\left(-\frac{\kappa^2}{8\pi}S_{\rm L}-S_{\text{cosm}}\right)\delta\left(A-\int\mathrm{d}^2x\sqrt{g_0}\,e^{2\sigma}\right)
\end{align*}
where $\kappa^2=\frac{26-c}{3}$. One of the main difficulties when computing the partition function lies in the complicated non-flat measure $\mathcal{D}\sigma$ for the conformal factor.

KPZ were the first to characterize this partition function at fixed area \cite{Knizhnik:1988ak} for a two-dimensional quantum gravity. They derived the scaling law
\begin{align*}
Z[A]\sim e^{-\mu_c^2A}A^{\gamma_{\rm str}-3}
\end{align*}
and a formula for the string susceptibility $\gamma_{\rm str}$, in the light-cone gauge for genus zero. Working in the conformal gauge, \cite{Distler:1988jt} and \cite{David:1988hj} managed to generalize the KPZ formula for surfaces of arbitrary genus, making several simplifying assumptions and using consistency conditions:
\begin{align*}
\gamma_{\rm str}=2+2\left(h-1\right)\frac{\sqrt{25-c}}{\sqrt{25-c}-\sqrt{1-c}}~.
\end{align*}
While alternative derivations such as \cite{Duplantier:2009np} and \cite{David:2008su} for $c\le 1$ and $h=0$ have more recently been obtained, no obvious way to circumvent the so-called ``$c=1$ barrier'', stating that this formula turns complex for $1<c<25$, has yet been found.

The recent development of efficient multi-loop regularization methods on curved space-times \cite{Bilal:2013iva} opened the way for a precise and well-defined perturbative computation of this fixed-area partition function in the K\"ahler formalism where the conformal factor is traded for the (Laplacian of the) K\"ahler potential as the basic quantum field. In  \cite{Bilal:2013ska} the string susceptibility was computed in this framework up to one loop for surfaces of arbitrary genus using a somewhat more general quantum gravity action including the Liouville and Mabuchi actions; the latter corresponds to possible couplings to non-conformal matter. Of course, for conformal matter only, the one-loop KPZ result was reproduced. This was to be expected since the non-trivial nature of the quantum gravity integration measure only shows up at two and higher loops. 

In \cite{Bilal:2014mla}  this computation was then extended to two loops with the Liouville action only. The regularized fixed-area partition function depends on the cut-off $\Lambda$ and the area $A$ through divergent terms of the form
$A\Lambda^2$, $\ln A\Lambda^2$,   $\left(\ln A\Lambda^2\right)^2$ and 
$A\Lambda^2\ln A\Lambda^2$. While the first term only contributes to the divergent cosmological constant (which can be adjusted by a corresponding local counterterm), and the coefficient of the second term determines  $\gamma_{\rm str}$, the third and fourth terms are unwanted, non-local divergences. Quite non-trivially, all contributions to the third term added up to zero\,! However, this  was not the case for the $A\Lambda^2\ln A\Lambda^2$ divergences which remained. As carefully argued in \cite{Bilal:2014mla} one can and must introduce local counterterms other than just the cosmological constant. Such local counterterms then also contribute, via one-loop diagrams, to the two-loop partition function. In particular, they can cancel the $A\Lambda^2\ln A\Lambda^2$ divergences, but they could not cancel any $\left(\ln A\Lambda^2\right)^2$ divergences. Happily, the latter cancelled among themselves without any need of counterterms.
The precise coefficients of the counterterms  were determined up to regulator-independent finite constants by requiring that the two-point function of the K\"ahler fields, or equivalently of $\langle e^{2\sigma} e^{2\sigma}\rangle$,  be finite and regulator-independent. Their contributions  to the partition function was exactly the one required to make it finite and regulator-independent. Yet, two finite ``renormalization'' constants -- on which the two-loop contribution to $\gamma_{\rm str}$ depends -- remained undetermined. By a locality argument, one of these renormalization constants was fixed, precisely to the value consistent with the KPZ value of $\gamma_{\rm str}$. However, the other renormalization constant had no particular reason to be fixed to the KPZ value, thus allowing a one-parameter family of quantization schemes which could eventually open the possibility  to go beyond the $c=1$ barrier.

The presence of this free parameter is intriguing and a natural question is  whether the structure of the \ct action introduced at two-loops is enough to cancel also the divergences at three (and higher) loops or whether new counterterms, with additional undetermined finite renormalization constants are required. This is  the motivation for the present paper. In section \ref{sec:diagrams}, the Liouville action, measure and (two-loop) \ct actions are expanded to the order relevant for the computation of the partition function at three loops. In particular, this leads to new vertices. Then the three-loop vacuum diagrams are enumerated. As could be expected, there is quite a large number of these diagrams. In section \ref{sec:divergence}, the allowed divergences are investigated in some detail and the leading divergence  $\sim A\Lambda^2\left(\ln A\Lambda^2\right)^2$ is fully computed with the result
\begin{align*}
\ln Z[A]\Big|_{3-\text{loop}}^{\text{leading div}} = &\ \frac{A\Lambda^2}{4\pi\kappa^4}\left(\ln A\Lambda^2\right)^2\Big(-26 R_1[\varphi] +132 R_2[\varphi]-216 R_3[\varphi] + 96 R_4[\varphi] \Big)~,
\end{align*}
where the $R_i[\varphi]$ are four different regulator dependent constants.\footnote{
In the general spectral cut-off regularization scheme used,  one introduces quite arbitrary regularization functions $\varphi$ and then $R_i[\varphi]=\int_0^\infty d\alpha_1 \ldots d\alpha_i \, \varphi(\alpha_1) \ldots \varphi(\alpha_i) \frac{1}{\alpha_1 + \ldots +\alpha_i}$.
}
Since this divergence does not cancel, new counterterms  are required. 

Section \ref{sec:ct} is dedicated to the discussion of such new counterterms that contribute via two-loop diagrams to the three-loop partition function, and in particular to the freedom to adjust them to cancel the divergences in the partition function. Of course, to really determine the coefficients of these counterterms  one needs to compute the three (and four)-point functions at one loop and the two-point function at two loops and to require them to be finite and regulator independent. (Actually, just as in \cite{Bilal:2014mla}, this would fix the diverging, as well as the finite regulator-dependent parts of the counterterm coefficients, but not certain finite ``renormalization constants''.) While this computation is beyond the scope of the present paper, it is already interesting to check if every divergence can be cancelled through the introduction of local counterterms. We call a counterterm local if it is a local expression in the K\"ahler field and if its coefficient is local.  In particular, a counterterm coefficient involving $\ln A\Lambda^2$ is not local.  However, coefficients proportional to $\frac{1}{A}$ are allowed in the first place, since such terms already naturally appear through the measure action. (It is interesting though to require their absence from the combined counterterm and measure action, a condition that we will refer to as the ``strong locality condition''.) At two loops, such local counterterms  could cancel all the two-loop divergences but $\left(\ln A\Lambda^2\right)^2$. Thus, for consistency, this divergence had to cancel by itself, which was the case, as already mentioned above. The same situation repeats itself at three loops where the counterterms generate exactly the necessary terms to cancel all divergences of the three-loop partition function, except for a $\left(\ln A\Lambda^2\right)^3$ divergence which, if present, cannot be cancelled by a local counterterm. This divergence is present in individual three-loop diagrams but we show that the different contributions cancel among themselves. This is an encouraging result meaning that all the non-local divergences appearing through the computation of the partition function may be offset by local counterterms. We end with a discussion of how many counterterm  parameters one expects to be fixed and how many free finite ``renormalization constants"  remain after imposing cancellation of the divergences, of any regulator dependence and requiring the strong locality condition.

\newpage
\section{Three-loop framework}
\label{sec:diagrams}

\subsection{The K\"ahler formalism}

In two dimensions, any metric $g$ on a compact Riemann surface may be written in the conformal gauge in terms of a reference metric $g_0$ and the conformal factor $\sigma$. Moreover, in two dimensions all the metrics are Kähler's, so that one can rewrite the metric in terms of the Kähler potential $\phi$ (and the background metric $g_0$):
\vskip-7.mm
\begin{align}
g=e^{2\sigma}g_0~~,~~e^{2\sigma}=\frac{A}{A_0}\left(1-\frac{1}{2}A_0\Delta_0\phi\right)
\label{eq:sigma_phi}
\end{align}
where $A$ and $A_0$ are the areas of the metrics $g$ and $g_0$ respectively and $\Delta_0$ denotes the Laplacian for the reference metric. 

Throughout this paper, we will consider the Liouville action,
\begin{align}
S_{\rm L}\left[\sigma\right]=\int \mathrm{d}^2x\sqrt{g_0}\left[\sigma\Delta_0\sigma
+R_0\sigma\right].
\label{eq:Liouville_0}
\end{align}
The classical saddle points of this action are the constant curvature metrics $g_*$ of arbitrary area $A$ and genus $h$. Thus, choosing the background metric $g_0$ to be a constant curvature metric of given area $A_0$, the Liouville action may be trivially rewritten in terms of $\sigma$ and the rescaled $g_*$, $\Delta_*$, $R_*$ as
\begin{align}
S_{\rm L}\left[\sigma\right]=\int \dx\left[\sigma\Delta_*\sigma+R_*\sigma\right]\ ,
\label{eq:Liouville}
\end{align}
\vskip-3.mm
\noindent
where
\begin{align}
g_*=\frac{A}{A_0}g_0~~,~~\Delta_*=\frac{A_0}{A}\Delta_0~~,~~R_*=\frac{A_0}{A}R_0=\frac{8\pi(1-h)}{A}~.
\label{eq:scaling_relations}
\end{align}
The field considered in the following will not be exactly the Kähler potential but rather
\begin{align}
\tilde{\phi}=\frac{\kappa}{8\sqrt{\pi}}A\Delta_*\phi
\label{eq:phi_tilde}
\end{align}
which appears naturally when writing both the Liouville and the measure actions in the Kähler formalism. The explicit introduction of the factor containing $\kappa$, where
\begin{align}
\kappa^2=\frac{26-c}{3} \ ,
\label{kappavalue}
\end{align}
 allows the loop-counting parameter $\frac{1}{\kappa^2}$ to appear clearly in the expansion of the action  performed later-on. Note that the relation \eqref{eq:sigma_phi} defines $A$ and $\tilde{\phi}$ uniquely for given $\sigma$.

In quantum gravity one needs to integrate over the space of metrics modulo diffeomorphisms. As emphasized in \cite{Bilal:2013ska,Bilal:2014mla}, the integration measure $\mathcal{D}\sigma$ over the conformal factor $\sigma$ is not the measure of a free field. Using the parametrization \eqref{eq:sigma_phi}  induces a non-trivial measure \cite{Bilal:2013ska,Bilal:2014mla}:
\begin{align}
\mathcal{D}\sigma=\frac{\mathrm{d}A}{\sqrt{A}}\left[\text{Det}'\left(1-\frac{4\sqrt{\pi}}{\kappa}\tilde{\phi}\right)^{-1}\right]^{1/2}\mathcal{D}_*\tilde{\phi}
\label{eq:integration_measure}
\end{align} 
where $\mathcal{D}_*\tilde{\phi}$ is the standard free field integration measure in the background metric $g_*$ deduced from the metric $\|\delta\tilde{\phi}\|^2_*=\int\dx\delta\tilde{\phi}^2$. The notation Det$'$ means that the zero-modes are not taken into account when computing the determinant, which is consistent with the fact that $\tilde{\phi}$ has no zero-mode. The measure $\mathcal{D}_*\tilde{\phi}$ can be expressed in the traditional way by expanding $\tilde{\phi}$ in eigenmodes of the Laplace operator $\Delta_*$. Choosing $0=d^*_0<d^*_1\le d^*_2\le \cdots$ to be the eigenvalues of $\Delta_*$ and $\psi_r$ its eigenfunctions, that are chosen to be real, then
\begin{align}
\tilde{\phi}=\sum\limits_{r>0}c_r\psi_r\ , \quad
\Delta_*\psi_r=d^*_r\psi_r\ , \quad
\int\mathrm{d}^2x\sqrt{g_*}\psi_r\psi_s=\delta_{rs} \ , 
\label{eq:phi_eigenfunction}
\end{align} 
and the measure is defined as
\begin{align}
\mathcal{D}_*\tilde{\phi}=\prod\limits_{r>0}\mathrm{d}c_r~.
\label{eq:measure_eigenfunction}
\end{align}

The study made in \cite{Bilal:2014mla} showed that the insertion of a \ct action was required for the finiteness of the two-point function. Therefore, the quantum gravity partition function at fixed area one considers as a starting point for the present work is
\begin{align}
Z[A]&=\frac{e^{-\mu_c^2A}}{\sqrt{A}}\int\mathcal{D}_*\tilde{\phi}\left[\text{Det}'\left(1-\frac{4\sqrt{\pi}}{\kappa}\tilde{\phi}\right)^{-1}\right]^{1/2}\text{exp}\left(-S_{\rm ct}-\frac{\kappa^2}{8\pi}S_{\rm L}\left[\sigma[A,\tilde{\phi}]\right]\right) \nonumber\\
&=\frac{e^{-\mu_c^2A}}{\sqrt{A}}\int\mathcal{D}_*\tilde{\phi}~\text{exp}\left(-S_{\rm measure}-S_{\rm ct}-\frac{\kappa^2}{8\pi}S_{\rm L}\left[\sigma[A,\tilde{\phi}]\right]\right) .
\label{eq:partition_function}
\end{align}
The measure action is thus defined as 
\begin{align}
S_{\rm measure}=-\frac{1}{2}\ln\left[\text{Det}'\left(1-\frac{4\sqrt{\pi}}{\kappa}\tilde{\phi}\right)^{-1}\right]=\frac{1}{2}\text{Tr}\ln\left(1-\frac{4\sqrt{\pi}}{\kappa}\tilde{\phi}\right)  .
\label{eq:measure_action}
\end{align}

\subsection{Three-loop expansions of the actions}
\label{sub:3loopexpact}

To compute  the partition function at three loops, one has to expand the Liouville, the measure and the two-loop \ct actions around the classical saddle points up to order $\kappa^{-4}$. The classical solutions $\sigma_{\rm cl}$ are simply the constants $e^{2\sigma_{\rm cl}}=\frac{A}{A_0}$. Hence, from \eqref{eq:sigma_phi},
\begin{align}
\sigma-\sigma_{\rm cl}=\frac{1}{2}\ln\left(1-\frac{4\sqrt{\pi}}{\kappa}\tilde{\phi}\right)~~,~~\sigma_{\rm cl}=\frac{1}{2}\ln\frac{A}{A_0}~.
\label{eq:sigma_classique}
\end{align}
$\sigma_{\rm cl}$ being a constant, it disappears from the Laplacian term in \eqref{eq:Liouville}. Moreover, the curvature term being linear, one has
\begin{align}
S_{\rm L}[\sigma]=S_{\rm L}[\sigma_{\rm cl}]+S_{\rm L}[\sigma-\sigma_{\rm cl}]=4\pi(1-h)\ln\frac{A}{A_0}+S_{\rm L}\left[\frac{1}{2}\ln\left(1-\frac{4\sqrt{\pi}}{\kappa}\tilde{\phi}\right)\right].
\label{eq:Liouville_expansion}
\end{align}
	
Expanding the logarithm  straightforwardly leads to the expansion of  the Liouville action as relevant for the 3-loop computation:	
\begin{align}
\frac{\kappa^2}{8\pi}S_{\rm L}\left[\sigma\right]=\frac{\kappa^2}{2}(1-h)\ln\frac{A}{A_0}+&\int\dx\frac{1}{2}\tilde{\phi}(\Delta_*-R_*)\tilde{\phi} \nonumber \\
+\int \dx &\left[\frac{\sqrt{4\pi}}{\kappa}\tilde{\phi}^2(\Delta_*-\frac{2}{3}R_*)\tilde{\phi}+\frac{2\pi}{\kappa^2}\tilde{\phi}^2\Delta_*\tilde{\phi}^2\right. \nonumber \\
&+\left.\frac{16\pi}{3\kappa^2}\tilde{\phi}^3\Delta_*\tilde{\phi}-\frac{4\pi}{\kappa^2}R_*\tilde{\phi}^4\right] \nonumber \\
+\int \dx&\left[
\frac{16\pi^{3/2}}{\kappa^3}\left[\tilde{\phi}^4\left(\Delta_*-\frac{4}{5}R_*\right)\tilde{\phi}+\frac{2}{3}\tilde{\phi}^3\Delta_*\tilde{\phi}^2\right]\right.\nonumber \\
&+\left.\frac{(8\pi)^2}{\kappa^4}\left[\frac{4}{5}\tilde{\phi}^5\left(\Delta_*-\frac{5}{6}R_*\right)\tilde{\phi}+\frac{1}{2}\tilde{\phi}^4\Delta_*\tilde{\phi}^2+\frac{2}{9}\tilde{\phi}^3\Delta_*\tilde{\phi}^3\right]+\mathcal{O}(\kappa^{-5})\right]. 
\label{eq:Liouville_expansion_3}
\end{align} 	
The first term in the first line gives the classical contribution while the second term of the first line yields the one-loop determinant studied in \cite{Bilal:2013ska}. This latter term also provides a standard propagator for the present three-loop investigation, namely $\tilde{G}(x,y)=\braket{x|\left(\Delta_*-R_*\right)^{-1}|y}'$, where the tilde on $G$ and the prime indicate that the zero-mode is excluded. The second and third lines provide the vertices relevant for the two-loop vacuum diagrams computed in \cite{Bilal:2014mla}. The last two lines yield the quintic and sextic vertices which appears only at three (or higher)-loop computations. Note that the propagator does not carry any factor of $\kappa$, while the vertices involve various powers of $\frac{1}{\kappa}$ in such a way that an $L$-loop diagram is acompanied by a factor $\frac{1}{\kappa^{2(L-1)}}$.
In particular, in \eqref{eq:Liouville_expansion_3} we have displayed all
the Liouville vertices contributing to  vacuum diagrams with up to three loops. They can be grouped as follows. Two quintic vertices	
\begin{align}
\vcenter{\hbox{\includegraphics[scale=0.95]{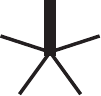}}}~=~-\frac{16\pi^{3/2}}{\kappa^3}\left(\Delta_*-\frac{4}{5}R_*\right)~,~~~
\vcenter{\hbox{\includegraphics[scale=0.95]{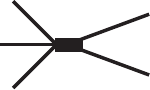}}}~=~-\frac{16\pi^{3/2}}{\kappa^3}\frac{2}{3}\Delta_*
\label{vertex_5}
\end{align}
and three sextic vertices
\begin{align}
\vcenter{\hbox{\includegraphics[scale=0.95]{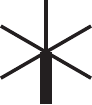}}}~=~-\frac{\left(8\pi\right)^2}{\kappa^4}\frac{4}{5}\left(\Delta_*-\frac{5}{6}R_*\right)~,~~~
\vcenter{\hbox{\includegraphics[scale=0.9]{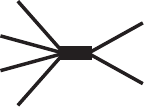}}}~=~-\frac{\left(8\pi\right)^2}{\kappa^4}\frac{1}{2}\Delta_*~,~~~
\vcenter{\hbox{\includegraphics[scale=0.9]{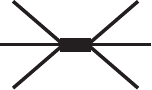}}}~=~-\frac{\left(8\pi\right)^2}{\kappa^4}\frac{2}{9}\Delta_*
\label{vertex_6}
\end{align}
for the ``pure three-loop'' contribution. The bold parts of the vertices encode the $\Delta_*$ acting on one or several propagators. For example, for the two quintic vertices, the $\left(\Delta_*-\frac{4}{5}R_*\right)$ in the first vertex acts on the single propagator connected to the bold line,  while in the second one $\Delta_*$ may act either on the product of the two propagators connected to the bold part of the vertex on the right or on the three other ones. The vertices already used to compute the two-loop vacuum diagrams in \cite{Bilal:2014mla} are one cubic and two quartic vertices:
\begin{align}
\vcenter{\hbox{\includegraphics[scale=0.95]{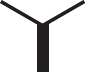}}}=~-\frac{\sqrt{4\pi}}{\kappa}\left(\Delta_*-\frac{2}{3}R_*\right)~,~~~
\vcenter{\hbox{\includegraphics[scale=0.95]{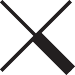}}}~=~-\frac{8\pi}{\kappa^2}\frac{2}{3}\Delta_*~,~~~
\vcenter{\hbox{\includegraphics[scale=0.95]{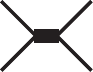}}}~=~-\frac{8\pi}{\kappa^2}\frac{1}{4}\left(\Delta_*-2R_*\right)~.
\label{vertex_4_3}
\end{align}
As it was already the case at two loops, the non-trivial measure action also contributes to the vacuum diagrams. To determine the expansion of the measure action \eqref{eq:measure_action} up to three loops, one needs to evaluate the trace of an operator $O$, which was done in \cite{Bilal:2014mla}: $\text{Tr}'O=\int \dx\sum\limits_{r>0}\psi_r^2(x)O(x)$. Hence,
\begin{align}
S_{\rm measure}=\int \dx\sum\limits_{r>0}\psi_r^2(x)\left(-\frac{\sqrt{4\pi}}{\kappa}\tilde{\phi}-\frac{4\pi}{\kappa^2}\tilde{\phi}^2-\frac{32\pi^{3/2}}{3\kappa^3}\tilde{\phi}^3-\frac{32\pi^2}{\kappa^4}\tilde{\phi}^4+\mathcal{O}(\kappa^{-5})\right).
\label{eq:measure_expansion}
\end{align} 
$\sum\limits_{r>0}\psi_r^2(x)$ is a formal writing which has to be regularized in a consistent way. After regularization, and since the considered metrics are of constant curvature, this quantity becomes independent of $x$. Since $\tilde{\phi}$ has no zero-mode, the first term in the action drops out. This action provides a quadratic, a cubic and a quartic vertex:
\begin{align}
\vcenter{\hbox{\includegraphics[scale=0.95]{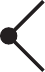}}}=~\frac{4\pi}{\kappa^2}\sum\limits_{r>0}\psi_r^2(x)~,~~
\vcenter{\hbox{\includegraphics[scale=0.95]{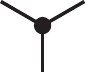}}}=~\frac{16\pi^{3/2}}{\kappa^3}\frac{2}{3}\sum\limits_{r>0}\psi_r^2(x)~,~~~
\vcenter{\hbox{\includegraphics[scale=0.95]{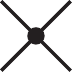}}}=~\frac{\left(8\pi\right)^2}{\kappa^4}\frac{1}{2}\sum\limits_{r>0}\psi_r^2(x)~.
\label{vertex_measure}
\end{align}
As already mentioned in section \ref{sec:intro}, counterterms  are required for the two-point function to be finite at one loop, as well as for the partition function to be finite at two loops \cite{Bilal:2014mla}. This two-loop \ct action is thus to be considered also for the three-loop computation:
\begin{align}
S_{\rm ct}=\frac{8\pi}{\kappa^2}\int \dx\left[\frac{c_\phi}{2}\tilde{\phi}(\Delta_*-R_*)\tilde{\phi}+\frac{c_R}{2}R_*\tilde{\phi}^2+\frac{c_m}{2}\tilde{\phi}^2\right]
\label{eq:ct_action}
\end{align}
where \cite{Bilal:2014mla}
\begin{align}
c_\phi(\Lambda,\alpha_i)&=\frac{1}{2\pi}\left[\frac{3}{2}\ln\frac{\alpha_2\alpha_3}{(\alpha_2+\alpha_3)^2}-1-\frac{2\alpha_2\alpha_3}{(\alpha_2+\alpha_3)^2}\right]+\widehat{c}_\phi~, \nonumber\\
c_R(\Lambda,\alpha_i)&=\frac{1}{2\pi}\left[\frac{3}{2}\ln\frac{\alpha_2\alpha_3}{(\alpha_2+\alpha_3)^2}-\frac{19}{12}-\frac{2\alpha_2\alpha_3}{(\alpha_2+\alpha_3)^2}\right]+\frac{\widehat{c}_R}{2\pi}~, \nonumber\\
c_m(\Lambda,\alpha_i)&=\frac{\Lambda^2}{2\pi}\left(\frac{2}{\alpha_2+\alpha_3}-\frac{5}{2\alpha_2}\right)+\frac{\widehat{c}_m}{A}~.
\label{eq:old_ct_values}
\end{align}
The $\widehat{c}_\phi$, $\widehat{c}_R$ and $\widehat{c}_m$ are regulator independent constants, while the other parts of these counterterms  are to be understood as $c[\varphi]=\int_0^\infty\mathrm{d}\alpha_2\mathrm{d}\alpha_3\varphi(\alpha_2)\varphi(\alpha_3)c(\alpha_2,\alpha_3)$. Note that they are local, as suitable for counterterms, except for $c_m$ because of the term $\frac{1}{A}$. Such a non-local term, however, naturally appears in the measure action \eqref{eq:measure_action_3}, making this \ct measure-like and hence acceptable. Moreover, imposing the ``strong locality condition'', i.e. locality on the joint measure and \ct action up to two loops fixed $\widehat{c}_m=-1$. This value of the \ct will be used in the following. The \ct action provides a quadratic vertex:
\begin{align}
\vcenter{\hbox{\includegraphics[scale=0.90]{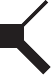}}}=~-\frac{4\pi}{\kappa^2}\left[c_\phi(\Delta_*-R_*)+(c_RR_*+c_m)\right]~.
\label{vertex_old_ct}
\end{align}

Note that all these vertices are normalized without including any symmetry factors so that one has to count all possible contractions when evaluating the diagrams. 

\subsection{Diagrams}
\label{sub:diagr}

We now enumerate all ``three-loop'' vacuum diagrams. More precisely, we give all diagrams contributing at order $\frac{1}{\kappa^4}$. This involves genuine three-loop diagrams made from the Liouville vertices only, as well as two-loop and one-loop diagrams involving also the vertices from the measure or counterterm action.
Combining all these vertices gives twenty-nine types of vacuum diagrams, each of them receiving contributions from subdiagrams. Fifteen of these diagrams come from pure Liouville contributions, nine involve the measure and six the two-loop counterterms . The decomposition of the diagrams is detailed hereafter.\\
The sextic vertices give one diagram, the ``flower diagram", which may be written as the sum of five subdiagrams:
\begin{align*}
\im{0.7}{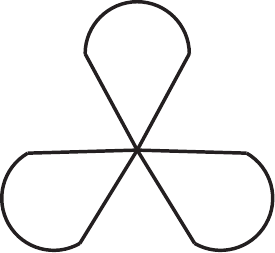}=&~15\im{0.8}{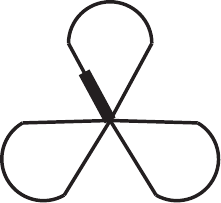}+~9\im{0.8}{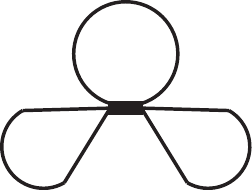}+~6~\im{0.85}{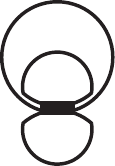}\\
&+~3\im{0.8}{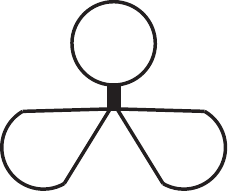}+~12~\im{0.9}{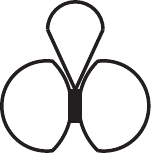}~~~.
\end{align*}
The weight factors in front of the different subdiagrams take into account the multiplicity of the diagram, including the symmetry factors and the contractions.
Combining the quintic and cubic vertices yields two types of diagram:
\begin{center}
\vspace{0.25cm}
\includegraphics[scale=0.7]{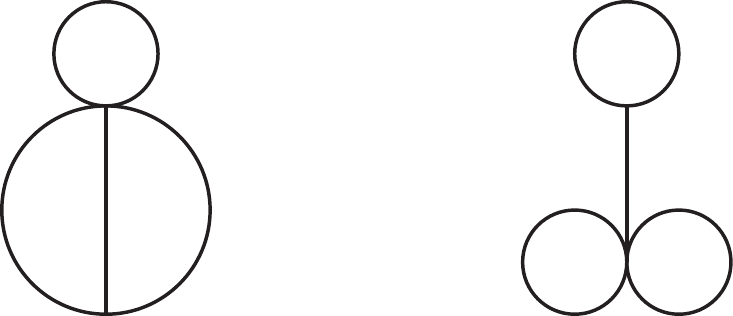}
\vspace{0.25cm}
\label{fig:diagrams_5_3}
\end{center}
composed of respectively eight and ten subdiagrams. Using two quartic vertices gives two diagrams:
\begin{center}
\vspace{0.25cm}
\includegraphics[scale=0.7]{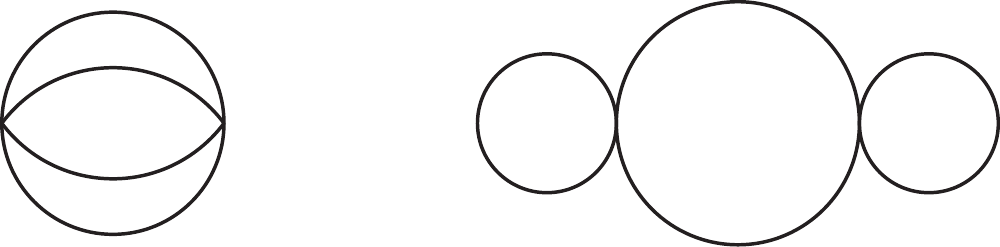}
\vspace{0.25cm}
\label{fig:diagrams_4_4}
\end{center}
made of respectively five and eleven subdiagrams. Five types of diagrams are built by a quartic vertex and two cubic vertices:
\begin{center}
\vspace{0.25cm}
\includegraphics[scale=0.7]{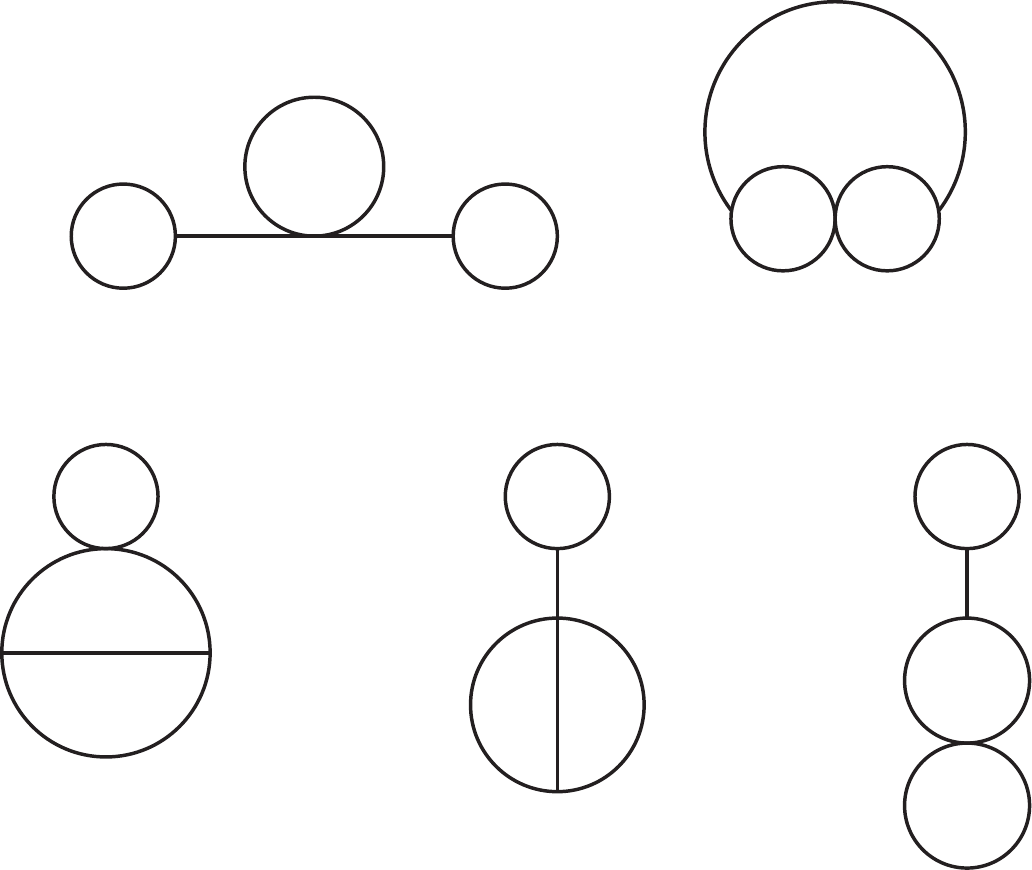}
\label{fig:diagrams_3_4_3}
\vspace{0.25cm}
\end{center}
These diagrams consist of thirteen subdiagrams each for the diagrams of the upper line, and of seventeen, ten and eighteen subdiagrams for the bottom line, from left to right. Finally, the last five pure Liouville diagrams come from using four cubic vertices:
\begin{center}
\vspace{0.25cm}
\includegraphics[scale=0.65]{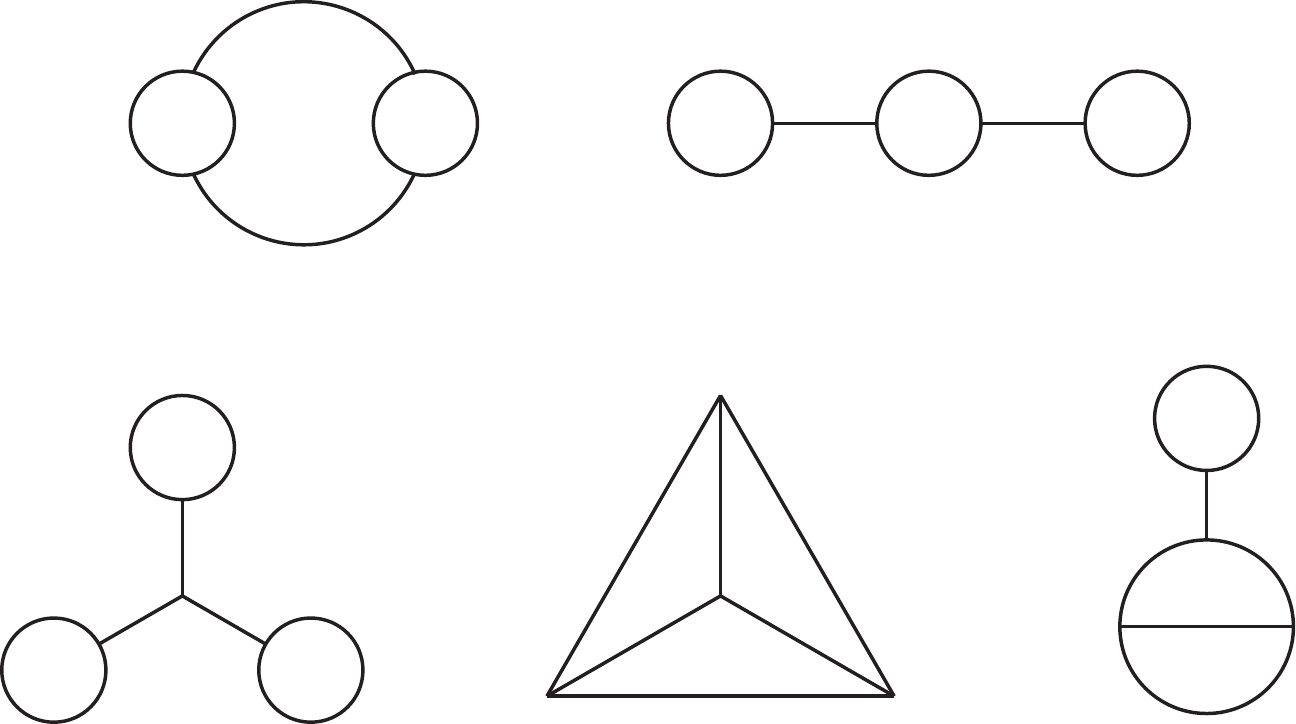}
\vspace{0.25cm}
\label{fig:diagrams_3_3_3_3}
\end{center}
composed of eleven, thirteen, six, six and eighteen subdiagrams, from left to right and from top to bottom.\\
The measure and counterterm  vertices contribute to fourteen diagrams. They may be classified according to the corresponding ``two-loop'' terminology. In \cite{Bilal:2014mla} there were four types of diagram: the ``figure-eight'', the ``setting sun'', the ``glasses'' and the ``measure'' diagrams. At the three-loop order, there are three ``figure-eight-like'' diagrams,
\begin{center}
\vspace{0.25cm}
\includegraphics[scale=0.6]{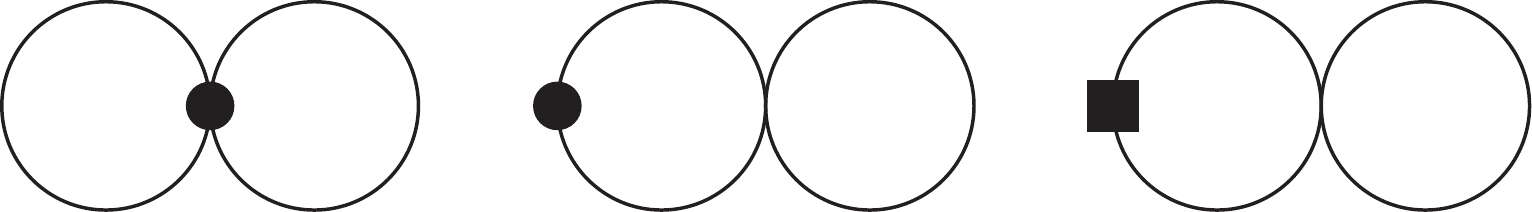}
\vspace{0.25cm}
\label{fig:diagrams_eight_like}
\end{center}
three ``setting sun-like'' diagrams,
\begin{center}
\vspace{0.25cm}
\includegraphics[scale=0.45]{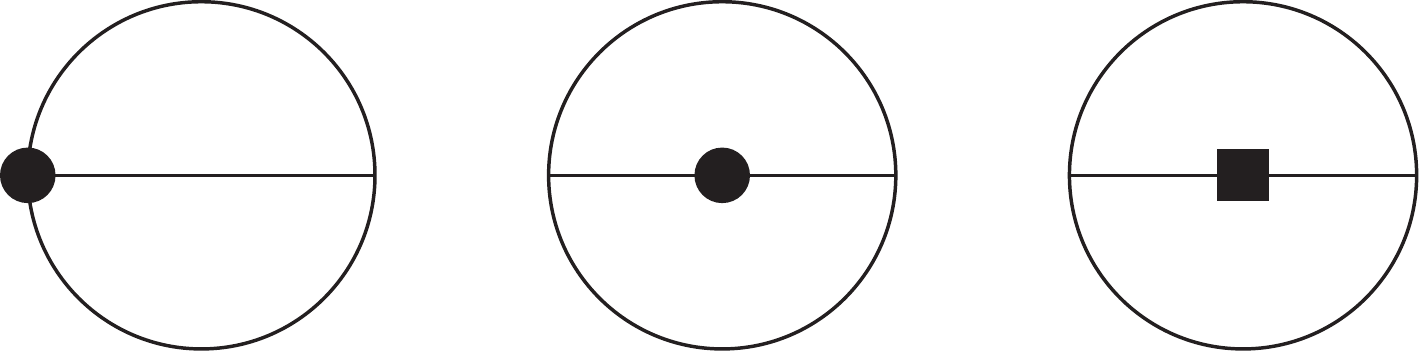}
\vspace{0.25cm}
\label{fig:diagrams_sun_like}
\end{center}
five ``glasses-like'' diagrams
\begin{center}
\vspace{0.25cm}
\includegraphics[scale=0.4]{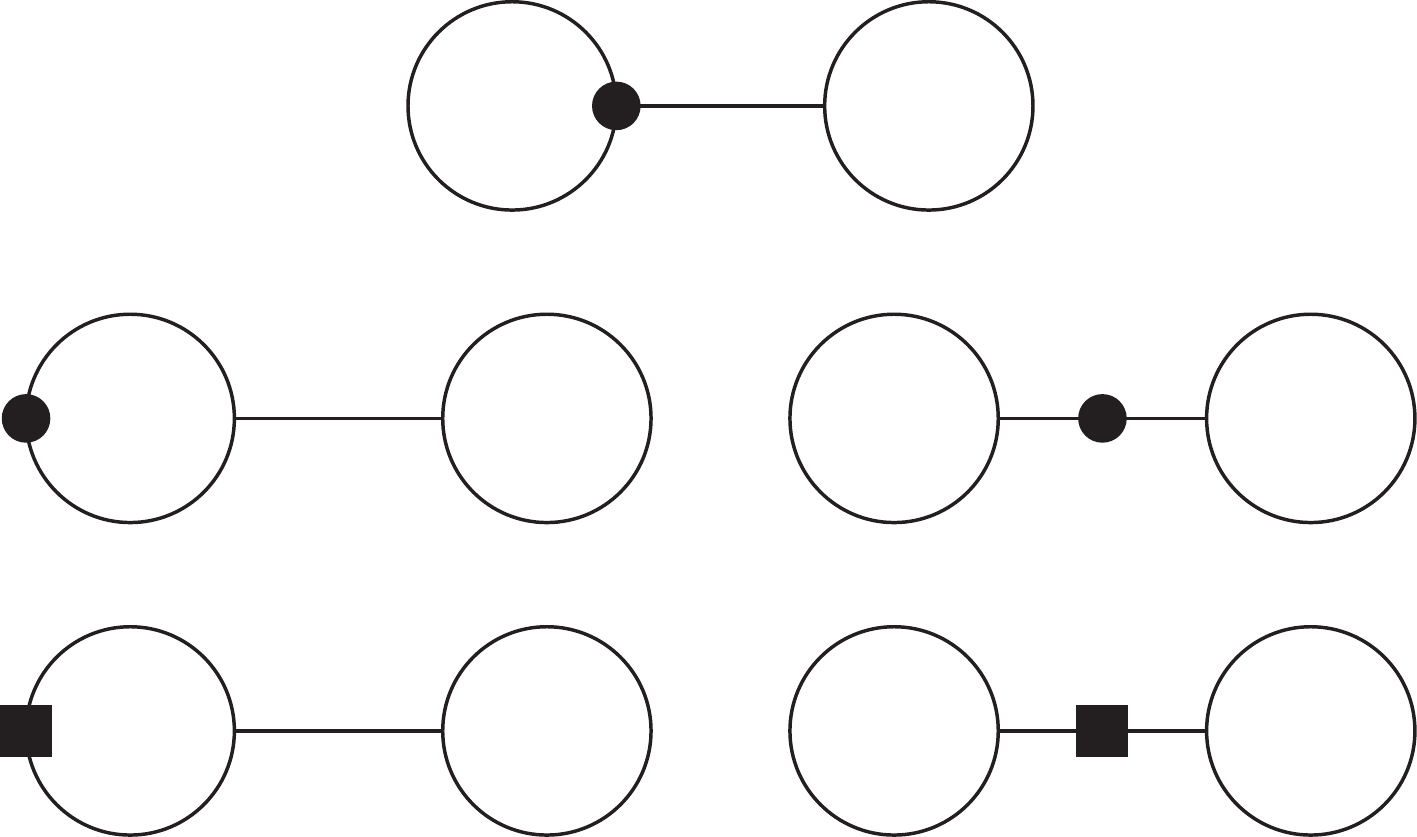}
\label{fig:diagrams_glasses_like}
\vspace{0.25cm}
\end{center}
and finally three ``measure-like'' diagrams.
\begin{center}
\vspace{0.25cm}
\includegraphics[scale=0.5]{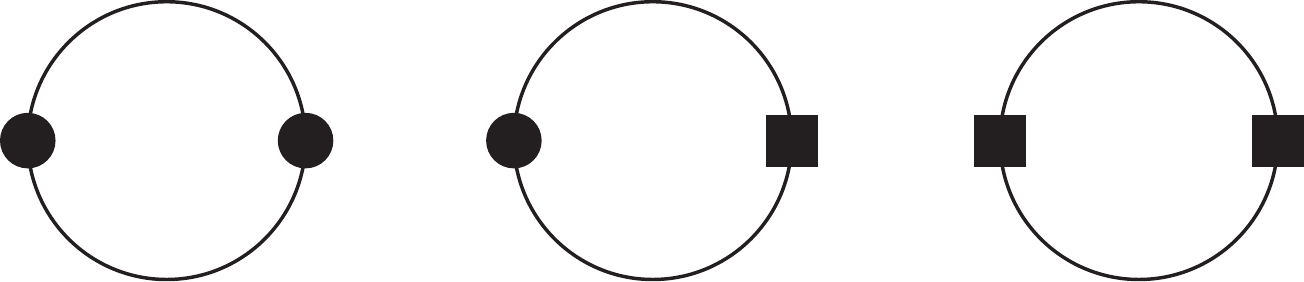}
\label{fig:diagrams_measure_like}
\vspace{0.25cm}
\end{center}
From left to right, both the ``figure-eight'' and ``setting sun'' diagrams have respectively one, four and five contributions. Concerning the ``glasses'' diagrams: the upper diagram has two contributions, and, from left to right, the diagrams in the second line have respectively four and three contributions, and the diagrams involving the \cts six and four respectively. Finally, the ``measure'' diagram on the right gets two contributions whereas both diagrams involving the measure vertex have no other subdiagram.

\subsection{Regularization}
\label{sub:regul}

The sums appearing in the diagrams, such as $\sum\limits_{r>0}\psi_r^2(x)$ encountered in the measure action or the Green's function $\tilde{G}(x,y)=\sum\limits_{r>0}\frac{\psi_r(x)\psi_r(y)}{\lambda_r}$, are formal writings of expressions which need to be regularized. The regularization scheme used in the present paper is the spectral cut-off approach developed in \cite{Bilal:2013iva}. This regularization scheme was used in the two-loop study of the partition function and details can be found in \cite{Bilal:2014mla}. The sums are regularized by inserting a rather arbitrary\footnote{
The function $\varphi$ must obey the obvious normalization condition $\int_0^\infty \mathrm{d}\alpha\varphi(\alpha)=1$, as well as certain regularity requirements at $0$ and $\infty$, but is otherwise arbitrary.
} 
regulator function $\varphi$ and a cut-off $\Lambda\rightarrow\infty$
\begin{align}
\tilde{G}(x,y)\rightarrow\int_0^\infty\mathrm{d}\alpha\varphi(\alpha)\sum\limits_{r>0}e^{-\frac{\alpha}{\Lambda^2}\lambda_r}\ \frac{\psi_r(x)\psi_r(y)}{\lambda_r}~,
\label{eq:reg_green}
\end{align}
$\lambda_r$ being the eigenvalues of the operator $D_*=\Delta_*-R_*$ appearing in the propagator. The tilde indicates that the zero-mode is excluded. The regularized quantities, and in particular the regularized Green's function, are related to the heat kernel or ``hatted heat kernel'' defined in \cite{Bilal:2013iva}:
\begin{align}
\kt(t,x,y)=&\sum\limits_{r>0}e^{-\lambda_r t}\, \psi_r(x)\psi_r(y)~, \nonumber\\
\hk(t,x,y)=&\int_t^\infty\mathrm{d}t'\kt(t',x,y)=\sum\limits_{r>0}\frac{e^{-\lambda_r t}}{\lambda_r}\, \psi_r(x)\psi_r(y)~.
\label{eq:def_k_hk}
\end{align}
These quantities satisfy the following relations:
\begin{align}
-\frac{\mathrm{d}}{\mathrm{d}t}\hk(t,x,y)=D_x\hk(t,x,y)=D_y\hk(t,x,y)=\kt(t,x,y)~.
\label{eq:eqheatkernel}
\end{align}
Furthermore, these sums are convergent for $t>0$, even for $x\rightarrow y$. For large $\Lambda$, $t=\frac{\alpha}{\Lambda^2}$ is small and the well-known small $t$-expansion of the heat kernel can be used, see \cite{Bilal:2013iva}:
\begin{align}
\kt(t,x,y)=&~\frac{e^{-\frac{l^2}{4t}}}{4\pi t}\left[a_0(x,y)+a_1(x,y)t+a_2(x,y)t^2+...\right]-\frac{e^{R_*t}}{A}~, \nonumber\\
\hk(t,x,y)=&~\tilde{G}(x,y)-\frac{1}{4\pi}\sum\limits_{k\geq 0}a_k(x,y)t^k\,E_{k+1}\Big(\frac{l^2}{4t}\Big)+\int_0^t\mathrm{d}t'\frac{e^{R_*t'}}{A}~.
\label{eq:k_hk_xy}
\end{align}
Due to the exponential term $e^{-l^2/4t}$, where $l$ is the geodesic distance between $x$ and $y$, this small $t$-expansion is also a short distance expansion and normal coordinates around $x$ or $y$ can be used. These expansions lead to the following expressions for the heat kernel $\kt$ and ``hatted heat kernel'' $\hk$ at coinciding points with the zero-modes excluded:
\begin{align}
\kt(t,x,x)=&\frac{1}{4\pi t}\left[1+\left(\frac{7}{6}R_*-\frac{4\pi}{A}\right)t+\left(\frac{41}{60}R_*-\frac{4\pi}{A}\right)R_*t^2\right]+\mathcal{O}(t^2)~, \nonumber\\
\hk(t,x,x)=&\frac{1}{4\pi}\left[-\ln \mu^2t+4\pi\tilde{G}_\zeta(x)-\gamma-\left(\frac{7}{6}R_*-\frac{4\pi}{A}\right)t\right]+\mathcal{O}(t^2)~.
\label{eq:k_xx}
\end{align}
Note that $R_*$ is a constant curvature and, hence, $\kt(t,x,x)$ does not depend on $x$. Furthermore, $\mu$ is an arbitrary scale and $\tilde{G}_\zeta(x)$ is the ``Green's function at coinciding points'', obtained  through a specific $\zeta$-function regularization scheme. It coincides, up to an additive constant, with the result obtained by subtracting the logarithmic short-distance singularity of $\tilde{G}(x,y)$ and by taking $y\rightarrow x$. Its area dependence is given by 
\begin{align}
\tilde{G}_\zeta^{A}=\tilde{G}_\zeta^{A_0}+\frac{1}{4\pi}\ln\frac{A}{A_0}~,
\label{eq:Gzeta}
\end{align}
such that $\hk$ may be rewritten as
\begin{align}
\hk(\frac{\alpha}{\Lambda^2},x,x)=&\tilde{G}_\zeta^{A_0}(x)+\frac{1}{4\pi}\left[\ln A\Lambda^2-\ln A_0\mu^2-\ln \alpha-\gamma-\left(\frac{7}{6}R_*-\frac{4\pi}{A}\right)\frac{\alpha}{\Lambda^2}\right]+\mathcal{O}(\Lambda^{-4})~.
\label{eq:hk_xx}
\end{align}
Despite the appearance and as explained in \cite{Bilal:2014mla}, the $\hk$ do not depend on the arbitrary $\mu$ and $A_0$ but only on $A\Lambda^2$, as well as on $\alpha$ and on various dimensionless moduli characterizing the geometry of the Riemann surface and coded in $\tilde{G}_\zeta$.

Furthermore, since the zero-modes are excluded from the sums, the following integrals vanish:
\begin{align}
\int \mathrm{d}^2x\sqrt{g_*(x)}\kt(t,x,y)=\int \mathrm{d}^2x\sqrt{g_*(x)}\hk(t,x,y)=0~.
\label{eq:vanish_k}
\end{align}
As an example, using  $\kt(t,x,x)$ to regularize $\sum_{r>0} \psi_r(x)^2$, the measure action becomes \begin{align}
S_{\rm measure}=\int \dx\frac{1}{4\pi}\left(\frac{\Lambda^2}{\alpha}+\frac{7}{6}R_*-\frac{4\pi}{A}+\mathcal{O}(\Lambda^{-2})\right)\left(-\frac{4\pi}{\kappa^2}\tilde{\phi}^2-\frac{32\pi^{3/2}}{3\kappa^3}\tilde{\phi}^3-\frac{32\pi^2}{\kappa^4}\tilde{\phi}^4+\mathcal{O}(\kappa^{-5})\right)~.
\label{eq:measure_action_3}
\end{align} 
This structure is very similar to those of the \ct action and in particular to those of the $c_m$ term \eqref{eq:old_ct_values}. 

In the sequel of this paper, when computing the regularized diagrams, all propagators are replaced as
\begin{align}
\widetilde G(x_i,y_i) \to \widetilde G_{\varphi}(x_i,y_i)=\int_0^\infty {\rm d}\alpha_i \,\varphi(\alpha_i)\, \hk\big( t_i=\frac{\alpha_i}{\Lambda^2},x_i,y_i\big) \ .
\label{Gregphi}
\end{align}
To simplify the notation, we will not write the $\int_0^\infty {\rm d}\alpha_1 \varphi(\alpha_1) \ldots \int_0^\infty {\rm d}\alpha_n \varphi(\alpha_n)$ and simply replace each propagator by $\hk(t_i,x_i,y_i)$ with the understanding that $t_i=\frac{\alpha_i}{\Lambda^2}\, $.

\section{On the divergences}
\label{sec:divergence}

\subsection{Expected divergence structure}
\label{sub:expdiv}

All  vacuum diagrams are dimensionless  and can depend on $A$ and $\Lambda$ only through the dimensionless combination $A\Lambda^2$. They contribute various divergences to the partition function. Standard power counting shows that any loop-diagram has a superficial degree of divergence equal to $2$. This means that divergences such as $A\Lambda^2\left(\ln A\Lambda^2\right)^\#$ are allowed. To have a more precise idea of the leading divergence, consider a diagram with $I$ internal lines and $V$ vertices. Each internal line, that is to say each regularized propagator $\hk$, gives a logarithmic divergence, according to \eqref{eq:hk_xx}. Besides, each vertex, carrying a Laplacian, transforms such a propagator into the corresponding heat kernel $\kt$ thanks to \eqref{eq:eqheatkernel}, leading to a quadratic divergence \eqref{eq:k_xx}. Each vertex also implies an integration over the manifold. Due to the term $e^{-l^2/4t}$ in the heat kernel \eqref{eq:k_hk_xy}, every integration contributes a factor $t_i\sim \frac{1}{~\Lambda^2}$ at most. (The subtraction of the zero-mode terms $\sim \frac{e^{R_*t}}{A}$ does not change the final conclusion.) For the last integration, however, all quantities to be integrated only depend on one point, hence no Gaussian integration can be performed and one just gets a factor of $A$. Putting everything together, the leading singularity of this $L$-loop vacuum diagram is
\begin{align}
\left(\ln A\Lambda^2\right)^{I-V}\left(\Lambda^2\right)^VA\left(\frac{1}{~\Lambda^2}\right)^{V-1}=\left(\ln A\Lambda^2\right)^{L-1}A\Lambda^2
\label{eq:counting_leading_div}
\end{align}
since $I-V=L-1$ for every diagram. Therefore, the leading divergence at three loops is $A\Lambda^2\left(\ln A\Lambda^2\right)^2$. 
Note that the vertices not only contain a Laplacian but also terms $\sim R_*\sim\frac{1}{A}$. Picking the contribution coming from $V-V'$ Laplacians and $V'$ terms $\sim R_*$ leads to the divergence
\begin{align}
\left(\ln A\Lambda^2\right)^{I-V+V'}\left(\Lambda^2\right)^{V-V'} A^{-V'}A\left(\frac{1}{~\Lambda^2}\right)^{V-1}=\left(\ln A\Lambda^2\right)^{L-1+V'}\left(A\Lambda^2\right)^{1-V'}~.
\label{eq:counting_subdiv}
\end{align}
For $V'>1$ this is vanishing. This means  that the  subleading divergence with the largest power of logarithms is $\left(\ln A\Lambda^2\right)^L$. Consequently, the expected divergences in $\ln Z[A]$ are
\begin{align}
\ln Z[A]\big|_{3-\text{loop}} = &d_1 A\Lambda^2\left(\ln A\Lambda^2\right)^2 + d_2 A\Lambda^2\ln A\Lambda^2 + d_3 A\Lambda^2 + d_4\left(\ln A\Lambda^2\right)^3 + d_5\left(\ln A\Lambda^2\right)^2 + d_6\ln A\Lambda^2 \nonumber\\
&+d_7+\mathcal{O}\left(\frac{\ln A\Lambda^2}{A\Lambda^2}\right).
\label{eq:z_divergences}
\end{align}
Note that the term $\sim\ln A\Lambda^2$, although divergent, has a physical meaning. Indeed, once all other divergences cancelled by appropriate counterterms, one has $\ln Z[A]\big|_{3-\text{loop}+\text{CT}}=\widetilde d_6\ln A\Lambda^2 
+\widetilde d_7+\mathcal{O}\left(\frac{\ln A\Lambda^2}{A\Lambda^2}\right)$ so that
\begin{align}
\lim_{\Lambda\to\infty}\ \frac{Z[A]}{Z[A_0]}\Big|_{3-\text{loop}+\text{CT}}=\left(\frac{A}{A_0}\right)^{\widetilde d_6} \ ,
\label{eq:gstrcontrib}
\end{align}
showing that $\widetilde d_6$ is the three-loop plus counterterm,  order $\frac{1}{\kappa^4}$, contribution to $\gamma_{\rm str}$.

\subsection{Cancellation of the $\Lambda^4$ divergence}
\label{sub:cancelL4}

Moreover, contrary to the preceeding, somewhat naive power counting argument, one observes ``unexpected'' $\Lambda^4$ divergences appearing in the diagrams indicated in \rtab{tab:N2_div}. They appear through the following integrals:
\begin{align}
J^{i}_j&=\int\dxy{x}{y}\widetilde{K}\left(t_i,x,x\right)\widetilde{K}(t_j,y,y)\widehat{\widetilde{K}}(t_m,x,y)\widehat{\widetilde{K}}(t_n,x,y)~, \nonumber\\
J^{i,j}_{k}&=\int\dxy{x}{y}\widetilde{K}\left(t_i+t_j,x,x\right)\widetilde{K}(t_k,y,y)\widehat{\widetilde{K}}(t_m,x,y)\widehat{\widetilde{K}}(t_n,x,y)~, \nonumber\\
J^{i,j}_{k,l}&=\int\dxy{x}{y}\widetilde{K}\left(t_i+t_j,x,x\right)\widetilde{K}(t_k+t_l,y,y)\widehat{\widetilde{K}}(t_m,x,y)\widehat{\widetilde{K}}(t_n,x,y)~,
\label{eq:N2_integrals}
\end{align} 
where $i$, $j$, $k$, $l$, $m$ and $n$ are different. From \eqref{eq:k_xx} one gets the leading divergences
\begin{align}
J^{i}_j\sim\frac{\Lambda^4}{\alpha_i\alpha_j}J~,~~~~~~J^{i,j}_{k}\sim\frac{\Lambda^4}{(\alpha_i+\alpha_j)\alpha_k}J~~~\text{and}~~~~J^{i,j}_{k,l}\sim\frac{\Lambda^4}{(\alpha_i+\alpha_j)(\alpha_k+\alpha_l)}J~,
\label{eq:N2_div}
\end{align}  
with $J=\int\dxy{x}{y}\widehat{\widetilde{K}}(t_m,x,y)\widehat{\widetilde{K}}(t_n,x,y)$. 
Thus these $\Lambda^4$ divergences come with three different structures in the $\alpha_i$. We display  all these unwanted divergences in  \rtab{tab:N2_div}. When summing them up, all three structures \eqref{eq:N2_div} cancel and there is no net $\Lambda^4$ divergence~!
\begin{table}[h]
\centering
\begin{tabular}{|c|c|c|c|c|c|c|c|c|c|c|c|}
 \hline
 & $\im{0.18}{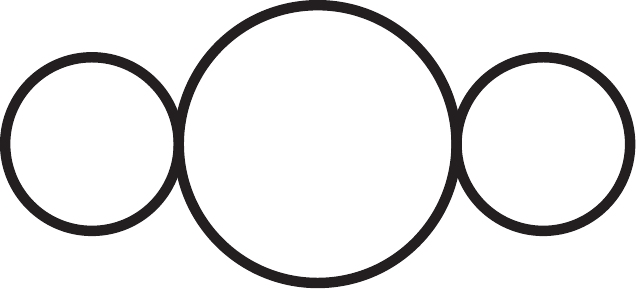}$ & $\im{0.25}{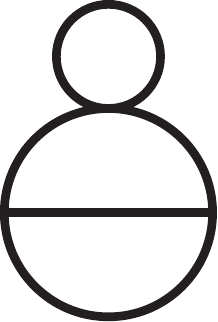}$ & $\im{0.25}{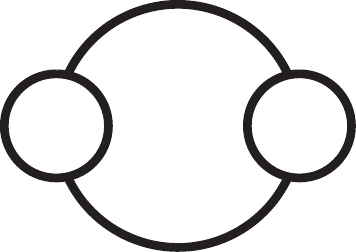}$ &  $\im{0.15}{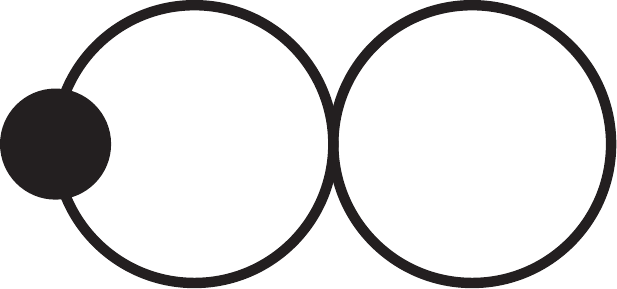}$ & $\im{0.15}{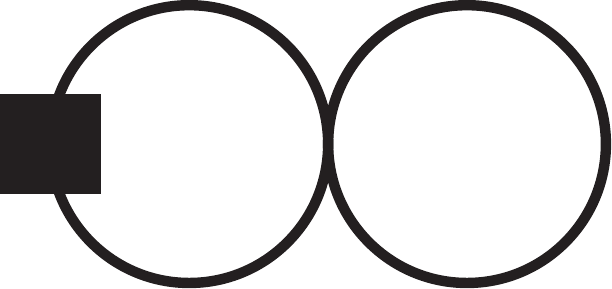}$ & $\im{0.18}{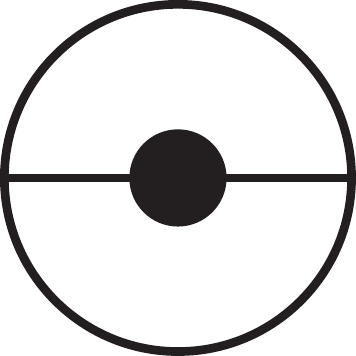}$ & $\im{0.18}{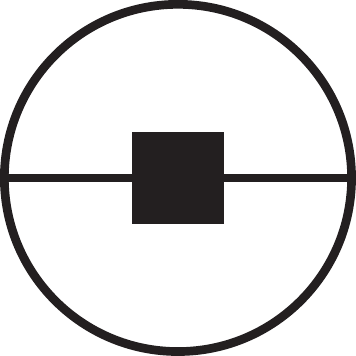}$ & $\im{0.21}{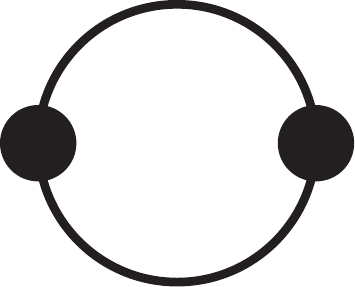}$ & $\im{0.21}{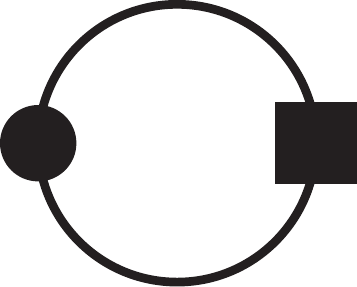}$ & $\im{0.21}{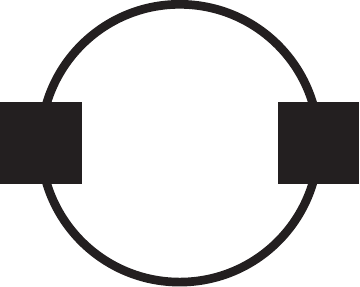}$ \rule[-0.4cm]{0pt}{1cm} & Total \\
 \hline
$J^{i}_{j}$    &9&   & &-3&-15& &  &$\frac{1}{4}$&$\frac{5}{2}$&$\frac{25}{4}$ & 0 \rule[-0.25cm]{0pt}{0.75cm}\\
  \hline
$J^{i+j}_{k}$  & &-12& &  & 12&2&10&             &     -2      &       -10     & 0 \rule[-0.2cm]{0pt}{0.7cm}\\
  \hline
$J^{i+j}_{k+l}$& &   &4&  &   & &-8&             &             &         4     & 0 \rule[-0.2cm]{0pt}{0.7cm}\\
 \hline
\end{tabular}
\caption{$\Lambda^4$ contributions from the diagrams}
\label{tab:N2_div}
\end{table}

\subsection{A simple computation: the flower diagram}
\label{sub:flower}

The true leading divergence	contributing to the partition function at three loops is in $A\Lambda^2\left(\ln A\Lambda^2\right)^2$. As already emphasised, the main goal of this work is to investigate this leading divergence, check that it does not ``miraculously'' cancel between the diagrams and  determine the structure of the required counterterms.

\begin{figure}[h]
\begin{center}
\includegraphics[scale=0.7]{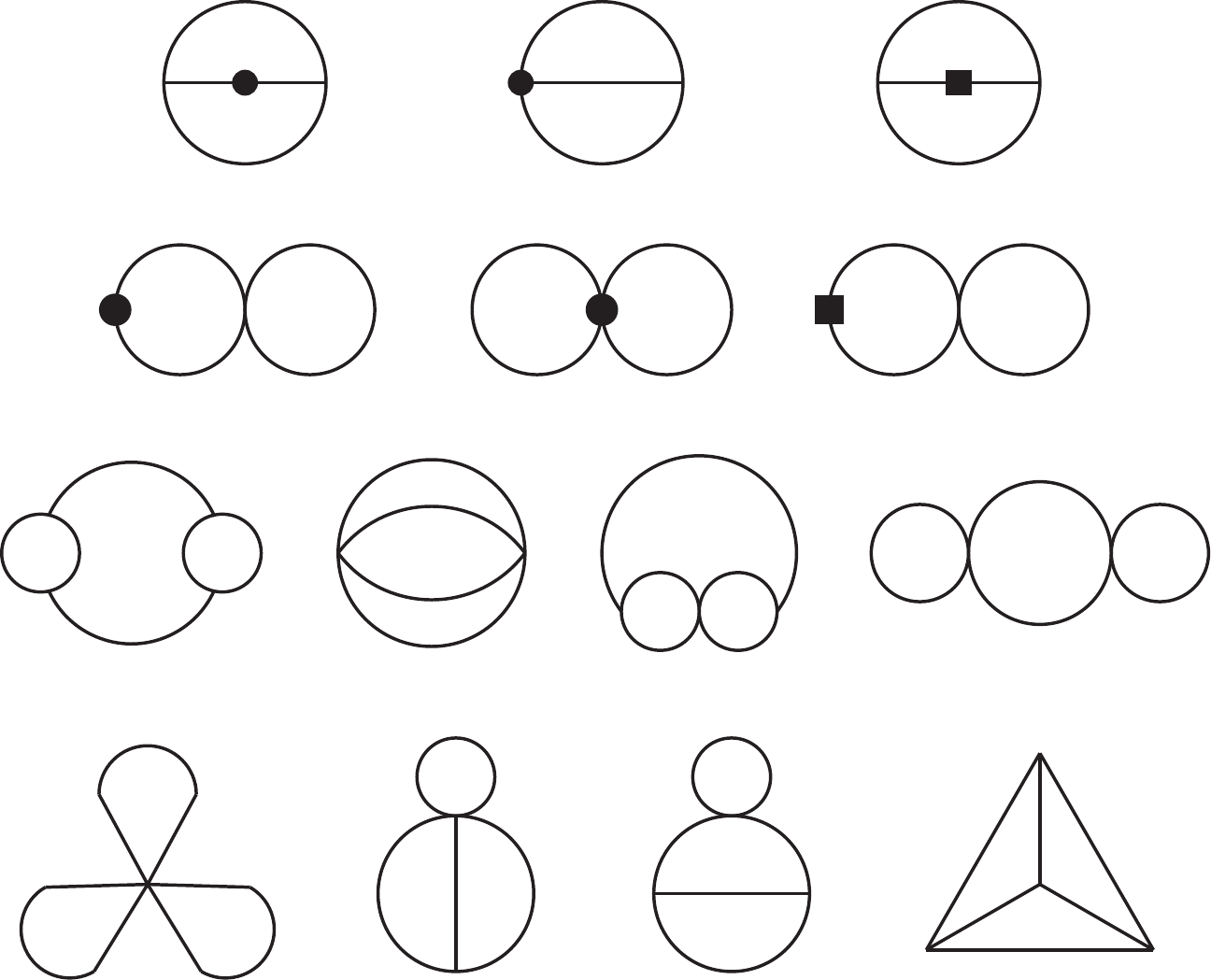}
\end{center}
\caption{Relevant diagrams for the leading divergence in $A\Lambda^2\left(\ln A\Lambda^2\right)^2$}
\label{fig:leading_diagrams}
\end{figure}

Out of the twenty-nine vacuum diagrams displayed in section~2.2, only the fourteen diagrams shown in \rfig{fig:leading_diagrams} contribute to the leading divergence in $A\Lambda^2\left(\ln A\Lambda^2\right)^2$. 
Note that all the diagrams with a single propagator between two vertices (i.e. one-particle reducible) do not contribute, as it was already the case in \cite{Bilal:2014mla}. This is because there is no zero-mode and a single  propagator connecting two parts of a vacuum diagram should carry only the zero-mode.\footnote{In flat space, by momentum conservation, such a propagator would carry zero momentum. In our curved geometry the argument is more complicated and such one-particle reducible diagrams can still be non-vanishing, but using \eqref{eq:vanish_k} one can show that they do not contribute to the present computation.}

Consider again the flower diagram made from the sextic vertices, whose decomposition in subdiagrams was given in  the previous section. Since only one vertex is involved, no integration has to be done to extract the divergences and it is the second simplest diagram to compute. (The simplest is the figure-eight diagram coming from the quartic measure vertex.) The first subdiagram may be written in our regularization as:
\begin{align}
I_{\hbox{\includegraphics[scale=0.15]{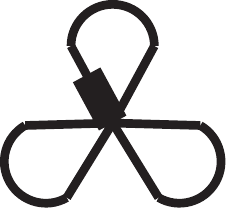}}}&=-\frac{4}{5} \frac{(8\pi)^2}{\kappa^4}\int\dx\hk\left(t_1,x,x\right)\hk\left(t_2,x,x\right)\left[\left(\Delta_*^x-\frac{5}{6}R_*\right)\hk(t_3,x,z)\right]_{x=z}\nonumber\\
&= - \frac{4}{5}\frac{(8\pi)^2}{\kappa^4}\int\dx\hk\left(t_1,x,x\right)\hk\left(t_2,x,x\right)\left(\kt(t_3,x,x)+\frac{1}{6}R_*\hk(t_3,x,x)\right)~,
\label{eq:fleur_diag1}
\end{align}
where \eqref{eq:eqheatkernel} was used. The second subdiagram is slightly more complicated, because of the Laplacian acting on several propagators:
\begin{align}
I_{\hbox{\includegraphics[scale=0.15]{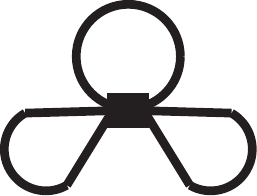}}}&=-\frac{2}{9} \frac{(8\pi)^2}{\kappa^4}\int\dx\hk\left(t_1,x,x\right)\left[\Delta_*^x\left(\hk\left(t_2,x,x\right)\hk(t_3,x,z)\right)\right]_{z=x}~.
\label{eq:fleur_diag2}
\end{align}
The Laplacian term gives
\begin{align}
\left[\Delta_*^x\left(\hk\left(t_2,x,x\right)\hk(t_3,x,z)\right)\right]_{z=x}=&~\hk\left(t_2,x,x\right)\left[\Delta_*^x\hk(t_3,x,z)\right]_{z=x}+\Delta_*^x\hk\left(t_2,x,x\right)\hk(t_3,x,x)\nonumber\\
&-2g^{ij}_*\partial^x_i\hk\left(t_2,x,x\right)\left[\partial^x_j\hk(t_3,x,z)\right]_{z=x}\nonumber\\
=&~\hk\left(t_2,x,x\right)\kt(t_3,x,x)+R_*\hk(t_2,x,x)\hk(t_3,x,x)\nonumber\\
&+\Delta_*^x\hk\left(t_2,x,x\right)\hk(t_3,x,x)-g^{ij}_*\partial^x_i\hk\left(t_2,x,x\right)\partial^x_j\hk(t_3,x,x)~.
\label{eq:laplacian_2_xx_xz}
\end{align}
Inserting \eqref{eq:laplacian_2_xx_xz} into \eqref{eq:fleur_diag2}, and integrating the last term by parts leads to:
\begin{align}
I_{\hbox{\includegraphics[scale=0.15]{gfleur2.pdf}}}&=-\frac{2}{9}\frac{(8\pi)^2}{\kappa^4}\int\dx\hk\left(t_1,x,x\right)\left[\hk\left(t_2,x,x\right)\kt(t_3,x,x)+R_*\hk\left(t_3,x,x\right)\hk(t_2,x,x)\right.\nonumber\\
&~~~~~~~~~~~~~~~~~~~~~~~~~~~~~~~~~~~~~~~~~~~~~\left.+~\frac{1}{2}~\hk\left(t_3,x,x\right)\Delta^x_*\hk(t_2,x,x)\right]~.
\label{eq:fleur_diag2_res}
\end{align}
Similarly, the third and fifth subdiagrams give
\begin{align}
I_{\hbox{\includegraphics[scale=0.2]{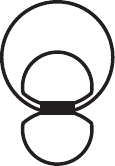}}}=&-\frac{2}{9} \frac{(8\pi)^2}{\kappa^4}\int\dx\left[\Delta^x_*\left(\hk(t_1,x,z)\hk(t_2,x,z)\hk(t_3,x,z)\right)\right]_{z=x}\nonumber\\
=&-\frac{2}{3}\frac{(8\pi)^2}{\kappa^4}\int\dx\hk\left(t_1,x,x\right)\left[\hk\left(t_2,x,x\right)\kt(t_3,x,x)+R_*\hk\left(t_3,x,x\right)\hk(t_2,x,x)\right.\nonumber\\
&~~~~~~~~~~~~~~~~~~~~~~~~~~~~~~~~~~~~~~~~~~~\left.-~\frac{1}{4}~\hk\left(t_3,x,x\right)\Delta^x_*\hk(t_2,x,x)\right]~,\nonumber\\
I_{\hbox{\includegraphics[scale=0.212]{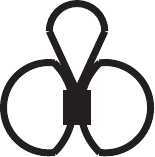}}}=&-\frac{1}{2} \frac{(8\pi)^2}{\kappa^4}\int\dx\hk(t_1,x,x)\left[\Delta^x_*\left(\hk(t_2,x,z)\hk(t_3,x,z)\right)\right]_{z=x}\nonumber\\
=&-\frac{(8\pi)^2}{\kappa^4}\int\dx\hk\left(t_1,x,x\right)\left[\hk\left(t_2,x,x\right)\kt(t_3,x,x)+R_*\hk\left(t_3,x,x\right)\hk(t_2,x,x)\right.\nonumber\\
&~~~~~~~~~~~~~~~~~~~~~~~~~~~~~~~~~~~~~~~~~~~\left.-~\frac{1}{8}~\hk\left(t_3,x,x\right)\Delta^x_*\hk(t_2,x,x)\right]~,
\label{eq:fleur_diag3_diag5}
\end{align}
while one reads directly the fourth subdiagram
\begin{align}
I_{\hbox{\includegraphics[scale=0.15]{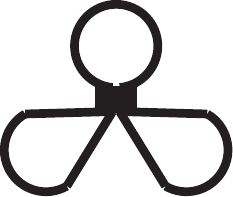}}}&=-\frac{1}{2} \frac{(8\pi)^2}{\kappa^4}\int\dx\hk(t_1,x,x)\hk(t_2,x,x)\Delta^x_*\hk(t_3,x,x)~.
\end{align}
The overall contribution from the flower diagram is thus
\begin{align}
I_{\hbox{\includegraphics[scale=0.125]{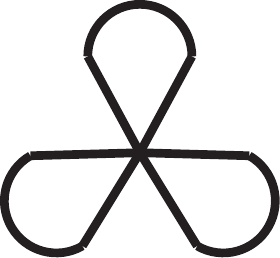}}}=&~15~I_{\hbox{\includegraphics[scale=0.15]{gfleur1.pdf}}}+9~I_{\hbox{\includegraphics[scale=0.15]{gfleur2.pdf}}}+6~I_{\hbox{\includegraphics[scale=0.2]{gfleur3.pdf}}}+3~I_{\hbox{\includegraphics[scale=0.15]{gfleur4.pdf}}}+12~I_{\hbox{\includegraphics[scale=0.212]{gfleur5.pdf}}}\nonumber\\
=&-\frac{(8\pi)^2}{\kappa^4}\int\dx\hk\left(t_1,x,x\right)\hk\left(t_2,x,x\right)\left[30\kt(t_3,x,x)+20R_*\hk(t_3,x,x)\right]~.
\label{eq:fleur_res}
\end{align}
(Note that the $\hk \hk \Delta\hk$ terms have cancelled.)
The leading divergence of the second term is in $\left(\ln A\Lambda^2\right)^3$. These divergences will be discussed in a seperate subsection where we show that all $\left(\ln A\Lambda^2\right)^3$ divergences cancel between the different diagrams.  The first term on the right-hand-side of  \eqref{eq:fleur_res} contributes to the leading divergence, giving  $-\frac{30}{\pi\kappa^4}A\Lambda^2\left(\ln A\Lambda^2\right)^2\int_0^\infty\mathrm{d}\alpha_3\frac{\varphi(\alpha_3)}{\alpha_3}$. 

\subsection{Leading divergence of the partition function per diagram}
\label{sub:leadingperdiag}

We have just seen that the contribution of the flower diagram to the leading divergence of the partition function is
\begin{align}
I_{\hbox{\includegraphics[scale=0.125]{gfleur.pdf}}}&=\frac{A\Lambda^2}{\pi\kappa^4}\left(\ln A\Lambda^2\right)^2\left[-\frac{30}{\alpha_1}\right]+\mathcal{O}(\Lambda^2\ln A\Lambda^2)~.
\label{eq:leading_fleur}
\end{align}
The only other diagram involving only one vertex is one of the measure ``figure-eight'' diagrams. It contributes
\begin{align}
I_{\hbox{\includegraphics[scale=0.09]{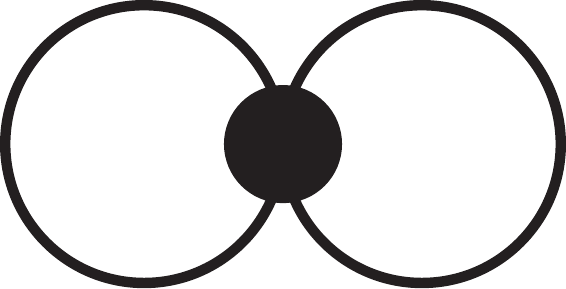}}}&=\frac{A\Lambda^2}{\pi\kappa^4}\left(\ln A\Lambda^2\right)^2\left[\frac{3}{2}\frac{1}{\alpha_1}\right]+\mathcal{O}(\Lambda^2\ln A\Lambda^2)~.
\label{eq:leading_huitm2}
\end{align}

There are six diagrams built from two vertices that contribute to the leading singularity: $~\im{0.22}{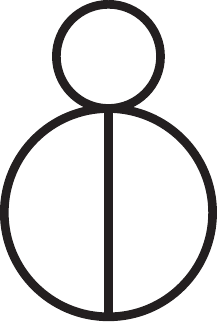}~$, $~\im{0.26}{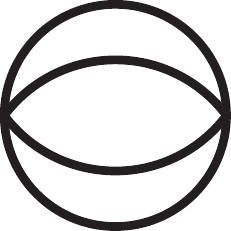}~$, $~\im{0.2}{gbonbon.pdf}~$, $~\im{0.17}{ghuitm1.pdf}~$, $\im{0.17}{ghuitct.pdf}~$ and $\im{0.15}{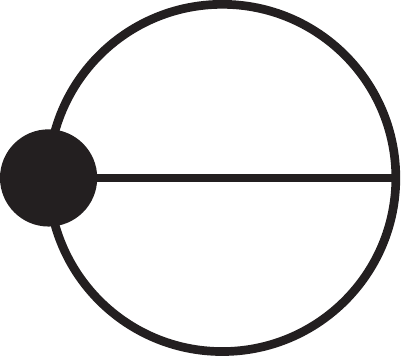}~$. The integrals to perform are similar to those done in \cite{Bilal:2014mla} to compute the two-loop vacuum diagrams. It is rather straightforward to obtain:
\begin{align}
I_{\hbox{\includegraphics[scale=0.125]{goeil.pdf}}}&=\frac{A\Lambda^2}{\pi\kappa^4}\left(\ln A\Lambda^2\right)^2\left[\frac{18}{\alpha_1+\alpha_2}\right]+\mathcal{O}(\Lambda^2\ln A\Lambda^2)~, \nonumber \\
I_{\hbox{\includegraphics[scale=0.08]{gbonbon.pdf}}}&=\frac{A\Lambda^2}{\pi\kappa^4}\left(\ln A\Lambda^2\right)^2\left[\frac{18}{\alpha_1}+\frac{9}{\alpha_1+\alpha_2}\right]+\mathcal{O}(\Lambda^2\ln A\Lambda^2)~, \nonumber \\
I_{\hbox{\includegraphics[scale=0.12]{gbonverti.pdf}}}&=\frac{A\Lambda^2}{\pi\kappa^4}\left(\ln A\Lambda^2\right)^2\left[\frac{24}{\alpha_1}+\frac{48}{\alpha_1+\alpha_2}\right]+\mathcal{O}(\Lambda^2\ln A\Lambda^2)~, \nonumber \\
I_{\hbox{\includegraphics[scale=0.09]{gsoleilm2.pdf}}}&=\frac{A\Lambda^2}{\pi\kappa^4}\left(\ln A\Lambda^2\right)^2\left[-\frac{2}{\alpha_1}\right]+\mathcal{O}(\Lambda^2\ln A\Lambda^2)~, \nonumber \\
I_{\hbox{\includegraphics[scale=0.09]{ghuitm1.pdf}}}&=\frac{A\Lambda^2}{\pi\kappa^4}\left(\ln A\Lambda^2\right)^2\left[-\frac{3}{\alpha_1}\right]+\mathcal{O}(\Lambda^2\ln A\Lambda^2)~, \nonumber \\
I_{\hbox{\includegraphics[scale=0.09]{ghuitct.pdf}}}&=\frac{A\Lambda^2}{\pi\kappa^4}\left(\ln A\Lambda^2\right)^2\left[-\frac{15}{\alpha_1}+\frac{12}{\alpha_1+\alpha_2}\right]+\mathcal{O}(\Lambda^2\ln A\Lambda^2)~.
\label{eq:leading_xy}
\end{align}
As always, according to our regularization scheme \eqref{eq:reg_green} and \eqref{Gregphi}, these expressions are to be understood as multiplied with the regulator functions $\prod_i\varphi(\alpha_i)$ and integrated $\prod_i\int_0^\infty {\rm d}\alpha_i$ . For instance $I_{\hbox{\includegraphics[scale=0.12]{gbonverti.pdf}}}$ contributes as $\frac{A\Lambda^2}{\pi\kappa^4}\left(\ln A\Lambda^2\right)^2c_{\hbox{\includegraphics[scale=0.09]{gbonverti.pdf}}}$, with $c_{\hbox{\includegraphics[scale=0.09]{gbonverti.pdf}}}=24\left(\int_0^\infty\mathrm{d}\alpha_1\frac{\varphi(\alpha_1)}{\alpha_1}+2\int_0^\infty\mathrm{d}\alpha_1\mathrm{d}\alpha_2\frac{\varphi(\alpha_1)\varphi(\alpha_2)}{\alpha_1+\alpha_2}\right)$ being a number once the regularization function $\varphi(\alpha)$ is chosen. 

Note that that the results for the diagrams involving the counterterm vertex, $I_{\hbox{\includegraphics[scale=0.09]{ghuitct.pdf}}}$ in \eqref{eq:leading_xy} and $I_{\hbox{\includegraphics[scale=0.09]{gsoleilct.pdf}}}$ in    \eqref{eq:leading_xyz_mct} below,  does not depend on the free (two-loop) renormalization constants $\widehat c_\phi$ and $\widehat c_R$ since the latter do not contribute to the leading divergence. In the next section we will carefully study the full contributions of the counterterms to all divergences and then, of course, the result will depend on $\widehat c_\phi$ and $\widehat c_R$.

When considering three vertices or more, computations become more technical. While for $~\im{0.15}{gsoleilm1.pdf}~$ and $~\im{0.15}{gsoleilct.pdf}~$ it is easy to get:
\begin{align}
I_{\hbox{\includegraphics[scale=0.09]{gsoleilm1.pdf}}}&=\frac{A\Lambda^2}{\pi\kappa^4}\left(\ln A\Lambda^2\right)^2\left[\frac{7}{2}\frac{1}{\alpha_1}\right]+\mathcal{O}(\Lambda^2\ln A\Lambda^2)~, \nonumber \\
I_{\hbox{\includegraphics[scale=0.09]{gsoleilct.pdf}}}&=\frac{A\Lambda^2}{\pi\kappa^4}\left(\ln A\Lambda^2\right)^2\left[\frac{35}{2}\frac{1}{\alpha_1}-\frac{14}{\alpha_1+\alpha_2}\right]+\mathcal{O}(\Lambda^2\ln A\Lambda^2)~,
\label{eq:leading_xyz_mct}
\end{align}
with $\im{0.22}{gbonneige.pdf}~$ and $\im{0.18}{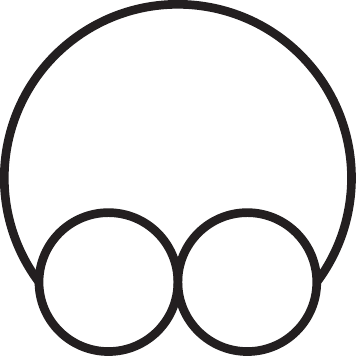}$ already, one stumbles over the same kind of technical difficulties as those faced when computing the one-loop two-point Green's function at coinciding points in \cite{Bilal:2014mla}.
One of the integrals encountered in $\im{0.18}{ggoggles.pdf}$ is for instance 
$$\int \mathrm{d}^2x\,\mathrm{d}^2y\,\mathrm{d}^2z\,\sqrt{g_*(x)g_*(y)g_*(z)}\, \hk(t_1,x,z)\kt(t_2,x,z)\hk(t_3,y,z)\kt(t_4,y,z)\kt(t_5,x,y)\ .$$ 
Trouble comes from the fact that the three $\kt$s in the integral force the three variables $x$, $y$ and $z$ to be all close to each other. For instance, integrating over $y$ through the term $\kt(t_4,y,z)$ requires to Taylor expand 
\begin{align}
\kt(t_5,x,y)=\kt(t_5,x,z)+(y-z)_i\partial_z^i\kt(t_5,x,z)+\frac{1}{2}(y-z)_i(y-z)_j\partial_z^i\partial_z^{j}\kt(t_5,x,z)+...
\label{eq:k_taylor}
\end{align}
When $x$, $y$ and $z$ are close, such that $l^2(x,y)\sim l^2(x,z)\sim \frac{1}{\Lambda^2}$, all terms in the expansion give contributions of the same order. One gets:
\begin{align}
\int \mathrm{d}^2x\mathrm{d}^2y\mathrm{d}^2z&\sqrt{g_*(x)g_*(y)g_*(z)}\hk(t_1,x,z)\kt(t_2,x,z)\hk(t_3,y,z)\kt(t_4,y,z)\kt(t_5,x,y)\nonumber\\
=&\int \mathrm{d}^2x\mathrm{d}^2z\sqrt{g_*(x)g_*(z)}\hk(t_1,x,z)\kt(t_2,x,z)\hk(t_3+t_4,z,z)\nonumber\\
&~~~~\times\left[\kt(t_5,x,z)-t_4\left(-\left.\frac{\mathrm{d}\kt(t,x,z)}{\mathrm{d}t}\right|_{t=t_5}+R_*\kt(t_5,x,z)\right)~\right.\nonumber\\
&~~~~~~\left.+\frac{t_4^2}{2}\left(\left.\frac{\mathrm{d^2}\kt(t,x,z)}{\mathrm{d}t^2}\right|_{t=t_5}-2R_*\left.\frac{\mathrm{d}\kt(t,x,z)}{\mathrm{d}t}\right|_{t=t_5}+R_*^2\kt(t_5,x,z)\right)+...\right] \nonumber\\ 
&-\frac{1}{A}\int \mathrm{d}^2x\mathrm{d}^2y\mathrm{d}^2z\sqrt{g_*(x)g_*(y)g_*(z)}\hk(t_1,x,z)\kt(t_2,x,z)\hk(t_3,y,z)\kt(t_5,x,y)~.
\label{eq:z11}
\end{align}
As just explained, the terms $+\ldots$ contribute at the same order and cannot be dropped.
Keeping only the terms that contribute to the leading singularity $A\Lambda^2\left(\ln A\Lambda^2\right)^2$, we have 
\begin{align}
\int \mathrm{d}^2x\mathrm{d}^2y\mathrm{d}^2z&\sqrt{g_*(x)g_*(y)g_*(z)}\hk(t_1,x,z)\kt(t_2,x,z)\hk(t_3,y,z)\kt(t_4,y,z)\kt(t_5,x,y)\nonumber\\
=\int &\mathrm{d}^2x\mathrm{d}^2z\sqrt{g_*(x)g_*(z)}\hk(t_1,x,z)\kt(t_2,x,z)\hk(t_3+t_4,z,z)\sum\limits_{n=0}^\infty\frac{t_4^n}{n!}\left.\frac{\mathrm{d}^n\kt(t,x,z)}{\mathrm{d}t^n}\right|_{t=t_5}.
\label{eq:z11_xz}
\end{align}
Furthermore, 
\begin{align}
\int \mathrm{d}^2x\sqrt{g_*(x)}\hk(t_1,x,z)\kt(t_2,x,z)\left.\frac{\mathrm{d}^n\kt(t,x,z)}{\mathrm{d}t^n}\right|_{t=t_5}=~\frac{\left(-1\right)^n n!}{\left(4\pi\right)^2}\left(\frac{\Lambda^2}{\alpha_2+\alpha_5}\right)^{n+1}\ln A\Lambda^2+\mathcal{O}(\Lambda^{2n+2})~,
\label{eq:hkkdnk}
\end{align}
so that one can easily resum all the terms. Therefore, the previous integral contributes to the leading divergence by:
\begin{align}
\frac{A\Lambda^2}{\left(4\pi\right)^3}\left(\ln A\Lambda^2\right)^2\sum\limits_{n=0}^\infty\frac{\left(-1\right)^n\alpha_4^n}{\left(\alpha_2+\alpha_5\right)^{n+1}}=\frac{A\Lambda^2}{\left(4\pi\right)^3}\left(\ln A\Lambda^2\right)^2\frac{1}{\alpha_2+\alpha_5+\alpha_4}~.
\label{eq:z11_contrib}
\end{align}
Of course, this is valid for $\frac{\alpha_4}{\alpha_2+\alpha_5}<1$. However, the initial expression was symmetric under exchange of $\alpha_2$ and $\alpha_4$ (upon also exchanging $\alpha_1$ and $\alpha_3$). Hence, if $\frac{\alpha_4}{\alpha_2+\alpha_5}>1$ one simply exchanges the roles of    $\alpha_2$ and $\alpha_4$ in the derivation (since now $\frac{\alpha_2}{\alpha_4+\alpha_5}<1$) and one gets the same result.

Considering carefully each integral, finally one gets\footnote{
Note again that the $\alpha_i$ are to be multiplied with $\varphi(\alpha_i)$ and integrated. This implies that any expression involving several $\alpha_i$ can be symmetrized and that one can also rename the indices. In particular, the $\frac{1}{\alpha_2+\alpha_5+\alpha_4}$ in \eqref{eq:z11_contrib} has been rewritten as $\frac{1}{\alpha_1+\alpha_2+\alpha_3}$.
} 
for $\im{0.22}{gbonneige.pdf}~$ and $~\im{0.18}{ggoggles.pdf}~$:
\begin{align}
I_{\hbox{\includegraphics[scale=0.12]{gbonneige.pdf}}}&=\frac{A\Lambda^2}{\pi\kappa^4}\left(\ln A\Lambda^2\right)^2\left[-\frac{21}{\alpha_1}-\frac{12}{\alpha_1+\alpha_2}-\frac{24}{\alpha_1+\alpha_2+\alpha_3}\right]+\mathcal{O}(\Lambda^2\ln A\Lambda^2)~, \nonumber \\
I_{\hbox{\includegraphics[scale=0.09]{ggoggles.pdf}}}&=\frac{A\Lambda^2}{\pi\kappa^4}\left(\ln A\Lambda^2\right)^2\left[-\frac{42}{\alpha_1+\alpha_2}-\frac{46}{\alpha_1+\alpha_2+\alpha_3}\right]+\mathcal{O}(\Lambda^2\ln A\Lambda^2)~.
\label{eq:leading_xyz_bng}
\end{align}

One encounters similar problems for the diagrams with four vertices  $\im{0.2}{gcylindre.pdf}$ and$~\im{0.16}{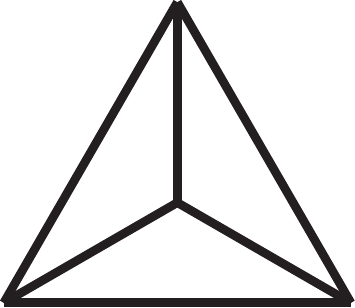}~$. Taylor expanding leads to series of divergent contributions. In addition to the series \eqref{eq:z11_contrib}, one obtains
\begin{align}
\sum\limits_{n=0}^\infty\sum\limits_{m=0}^\infty\binom{n+m}{n} \left(-1\right)^{n+m}\frac{\alpha_1^n\alpha_2^m}{\left(\alpha_3+\alpha_4\right)^{n+m+1}}=\frac{1}{\alpha_1+\alpha_2+\alpha_3+\alpha_4}~.
\label{eq:serie2}
\end{align}
More details on the integrals generating such series are given in the appendix. Thus, one gets:
\begin{align}
I_{\hbox{\includegraphics[scale=0.1]{gcylindre.pdf}}}&=\frac{A\Lambda^2}{\pi\kappa^4}\left(\ln A\Lambda^2\right)^2\left[\frac{14}{\alpha_1+\alpha_2}+\frac{16}{\alpha_1+\alpha_2+\alpha_3+\alpha_4}\right]+\mathcal{O}(\Lambda^2\ln A\Lambda^2)~, \nonumber \\
I_{\hbox{\includegraphics[scale=0.1]{gtetra.pdf}}}&=\frac{A\Lambda^2}{\pi\kappa^4}\left(\ln A\Lambda^2\right)^2\left[\frac{16}{\alpha_1+\alpha_2+\alpha_3}+\frac{8}{\alpha_1+\alpha_2+\alpha_3+\alpha_4}\right]+\mathcal{O}(\Lambda^2\ln A\Lambda^2)~.
\label{eq:leading_wxyz}
\end{align} 

Looking at \eqref{eq:leading_huitm2}, \eqref{eq:leading_xy} and \eqref{eq:leading_xyz_mct} one observes that the total leading contribution coming from the measure vanishes. Note that this was not the case for the two-loop contribution.\\
	
Adding the contributions of all the vacuum diagrams, \eqref{eq:leading_fleur}, \eqref{eq:leading_huitm2}, \eqref{eq:leading_xy}, \eqref{eq:leading_xyz_mct}, \eqref{eq:leading_xyz_bng} and \eqref{eq:leading_wxyz}, one gets the coefficient $d_1$ of $A\Lambda^2\left(\ln A\Lambda^2\right)^2$ in the logarithm of the partition function, cf \eqref{eq:z_divergences}:
\begin{align}
d_1=\frac{1}{4\pi\kappa^4}\left[-\frac{26}{\alpha_1}+\frac{132}{\alpha_1+\alpha_2}-\frac{216}{\alpha_1+\alpha_2+\alpha_3}+\frac{96}{\alpha_1+\alpha_2+\alpha_3+\alpha_4}\right]~.
\label{eq:result_leading}
\end{align}
We see that the leading divergence in $A\Lambda^2\left(\ln A\Lambda^2\right)^2$ is not vanishing and new counterterms  will be required. They should be determined by ensuring that the one-loop three-point and four-point  functions, as well as the two-loop two-point function be all finite. The computation of these one-loop $n$-point functions is beyond the scope of this paper, but it is nevertheless already interesting to look at the possible counterterms  one could consider and to calculate their contributions to the various divergences of the partition function. This will be done in the next section.

\subsection{Cancellation of the $\left(\ln A\Lambda^2\right)^3$ divergence}

Below, when we compute the counterterm contributions to the three-loop partition function, we will see that local counterterms with local coefficients (i.e.~not involving explicitly $\ln A\Lambda^2$) cannot give contributions to the  $\left(\ln A\Lambda^2\right)^3$ divergence. Now, it is easy to see that such 
$\left(\ln A\Lambda^2\right)^3$ divergences are present in individual three-loop diagrams.  In particular, this was the case for the flower diagram, see \eqref{eq:fleur_res} and the remarks that followed.
The only way to ensure finiteness of the partition function then is that these individual divergences cancel between the three-loop vacuum diagrams. Among the twenty-nine diagrams, eight contribute to the $\left(\ln A\Lambda^2\right)^3$ divergence. Their contributions are not too difficult to compute. We display the result in \rtab{tab:ln3_diagrams}. 
Indeed, when summed, they vanish! This is similar to what happened for the $\left(\ln A\Lambda^2\right)^2$ divergence in the two-loop partition function, and one expects the $\left(\ln A\Lambda^2\right)^L$ divergence to cancel  in the $L$-loop partition function.

\begin{table}[h]
\centering
\begin{tabular}{|c|c|c|c|c|c|c|c|c|c|}
 \hline
 & $\im{0.275}{gfleur.pdf}$ & $\im{0.29}{goeil.pdf}$ & $\im{0.18}{gbonbon.pdf}$
 & $\im{0.25}{gbonverti.pdf}$ & $\im{0.25}{gbonneige.pdf}$ & $\im{0.2}{ggoggles.pdf}$ & $\im{0.25}{gcylindre.pdf}$ & $\im{0.21}{gtetra.pdf}$ & Total \rule[-0.4cm]{0pt}{1cm} \\
 \hline
$\frac{8(1-h)}{\pi\kappa^4}\left(\ln A\Lambda^2\right)^3$&-20&18&18&60&-42&-86&26+$\frac{4}{3}$&24+$\frac{2}{3}$ & 0 \rule[-0.25cm]{0pt}{0.75cm}\\
  \hline
\end{tabular}
\caption{$\left(\ln A\Lambda^2\right)^3$ contributions from the diagrams}
\label{tab:ln3_diagrams}
\end{table}

\newpage

\section{Counterterms}
\label{sec:ct}

There are several types of counterterms  one may add in the three-loop computation. Cubic or quartic counterterms  lead to diagrams similar to the ones generated by the cubic and quartic measure vertices. One may also expand the coefficients of the quadratic counterterms  already present in the two-loop computation of  \cite{Bilal:2014mla} and consider their $\kappa^{-4}$ contributions. Of course, only local counterterms will be introduced. This means, on the one hand, that the counterterms are polynomial in the K\"ahler field $\tilde\phi$ with only finitely many derivatives acting on them, and, on the other hand, that the coefficients of these counterterms are local expressions. In particular, a counterterm coefficient involving the area e.g.~through $\ln A \Lambda^2$ is non-local. However, following \cite{Bilal:2014mla},  we do allow for counterterm coefficients $\sim\frac{1}{A}$ since they are already present in the measure action due to the absence of the zero-mode. Remarkably, imposing a ``strong locality condition'', i.e. absence of these $\frac{1}{A}$ terms, on the joint measure and \ct action of the two-loop computation \cite{Bilal:2014mla} fixed one of the two finite renormalization constants (namely $\widehat{c}_m$) precisely to the KPZ value. In this section, we will write out the counterterms  contributing to the partition function at the same order as the three-loop diagrams, i.e.~at order $\frac{1}{\kappa^4}$ and give their diverging contributions to $\ln Z[A]$. Since the divergences in $A\Lambda^2$ can always be absorbed in the cosmological constant they will be ignored in the following. Similarly, we will not spell out the finite contributions of the counterterms.

\subsection{Cubic conterterms}
\label{sub:cubicCT}

The new counterterms  one may introduce are cubic and quartic ones. The allowed cubic \ct action is
\begin{align}
S_{\rm ct}^c=\frac{16\pi^{3/2}}{\kappa^3}\frac{1}{2}\int\dx \left[f_\phi\tilde{\phi}^2(\Delta_*-R_*)\tilde{\phi}+f_R R_*\tilde{\phi}^3+f_m\tilde{\phi}^3\right]
\label{eq:ct_cubic_action}
\end{align}
where
\begin{align}
f_\phi&=f_\phi^{(1)}~, \nonumber\\
f_R&=f_R^{(1)}~, \nonumber\\
f_m&=f_m^{(1)}\Lambda^2+\frac{f_m^{(2)}}{A}~.
\label{eq:ct_cubic_values}
\end{align}
By dimensional analysis, the coefficients $f_i^{(1)}$  and $f_i^{(2)}$ are dimensionless ``numbers". As already emphasized in the two-loop analysis of \cite{Bilal:2014mla} they may depend on the regularization through the $\alpha_i$ and are then to be integrated with the given $\varphi(\alpha_i)$, resulting in a number. But they do not depend on the cut-off $\Lambda^2$. 
The action \eqref{eq:ct_cubic_action}
contributes via the two two-loop diagrams $~\im{0.25}{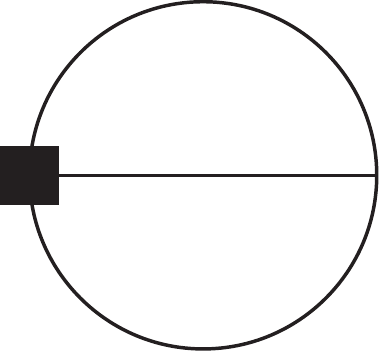}~$ and $~\im{0.3}{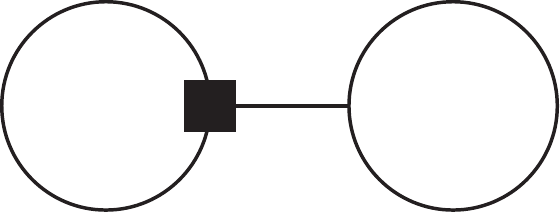}~$ at the same order in $\kappa^{-4}$ as the three-loop diagrams studied above.\\

We first show that the glasses diagram $~\im{0.3}{new_ct_glasses.pdf}~$ gives no relevant contribution. It may be written as a sum of four subdiagrams. One gets:
\begin{align}
I_{\hbox{\includegraphics[scale=0.12]{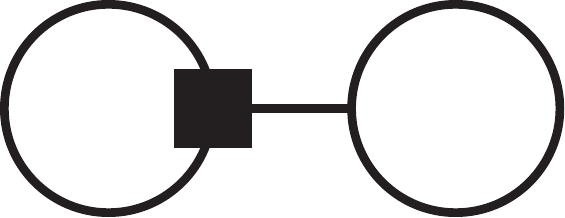}}}~=~\frac{1}{4}\frac{\left(8\pi\right)^2}{\kappa^4}\int&\dxy{x}{y}\hk(t_1,x,x)\hk(t_2,y,y)\nonumber\\
&\left\lbrace f_\phi\left(-\frac{\mathrm{d}\widetilde{K}(t_,x,y)}{\mathrm{d}t}\right)_{t=t_3}+\left[f_\phi R_*+3\left(f_m+f_R R_*\right)\right]\kt(t_3,x,y)\right.\nonumber\\
&~~~~~~~~\left.+3R_*\left(f_m+f_R R_*\right)\hk(t_3,x,y)\right\rbrace~.
\label{eq:glasses_ct_diagram}
\end{align} 
Integrating and taking into account the absence of zero-modes leads to:
\begin{align}
I_{\hbox{\includegraphics[scale=0.12]{gglassesnewct.pdf}}}~=~\frac{\left(4\pi\right)^2}{\kappa^4}&\left\lbrace f_\phi\int\dx ~\tilde{G}_\zeta\Delta_*\tilde{G}_\zeta\right. \nonumber\\
&+3\left(f_m+f_RR_*\right)\left(\int \dx ~\tilde{G}_\zeta(x)^2-\frac{1}{A}\int \mathrm{d}^2x\mathrm{d}^2y\sqrt{g_*(x)g_*(y)}~\tilde{G}_\zeta(x)\tilde{G}_\zeta(y)\right)\nonumber\\
&\left.+3\left(f_m+f_RR_*\right)R_*\int\mathrm{d}^2x\mathrm{d}^2y\sqrt{g_*(x)g_*(y)}~\tilde{G}_\zeta(x)\hk(t_3,x,y)\tilde{G}_\zeta(y)\right\rbrace~.
\label{eq:glasses_ct_res}
\end{align} 
Using the scaling relation \eqref{eq:scaling_relations} and \eqref{eq:Gzeta}, one may rewrite this as
\begin{align}
I_{\hbox{\includegraphics[scale=0.12]{gglassesnewct.pdf}}}~=~\frac{\left(4\pi\right)^2}{\kappa^4}&\left\lbrace f_\phi\int\mathrm{d}^2x\sqrt{g_0} ~\tilde{G}_\zeta^{A_0}\Delta_0\tilde{G}_\zeta^{A_0}\right. \nonumber\\
&+3\left(\frac{A}{A_0} f_m+f_R R_0\right)\left(\int \mathrm{d}^2x\sqrt{g_0}~\tilde{G}_\zeta^{A_0}(x)^2-\frac{1}{A_0}\int\mathrm{d}^2x\mathrm{d}^2y\sqrt{g_0(x)g_0(y)}~\tilde{G}_\zeta^{A_0}(x)\tilde{G}_\zeta^{A_0}(y)\right)\nonumber\\
&\left.+3\left(\frac{A}{A_0} f_m+f_RR_0\right)R_0\int\mathrm{d}^2x\mathrm{d}^2y\sqrt{g_0(x)g_0(y)}~\tilde{G}_\zeta^{A_0}(x)\hk_0(\frac{A_0}{A} t_3,x,y)\tilde{G}_\zeta^{A_0}(y)\right\rbrace~.
\label{eq:glasses_ct_A0}
\end{align} 
The first term is obviously independent of the area $A$ and thus of no interest here. The only $A$ dependence in the second line comes from the $\frac{A}{A_0} f_m$ term through $\frac{f_m^{(1)}}{A_0}A\Lambda^2$. However, the parenthesis being $A$ independent, this term can be included in the cosmological constant and is not significant. The last term is slightly more subtle to handle because of the remaining $\hk_0(\frac{A_0}{A} t_3,x,y)$ term. For the non divergent counterterms $f_R^{(1)}$ and $\frac{f_m^{(2)}}{A}$, the short-distance logarithmic singularity  in $\hk_0(\frac{A_0}{A} t_3,x,y)$ being integrable, one may take the limit $t_3\rightarrow\infty$. Doing so leads to an $A$ independent quantity. Finally, doing a finite expansion in $x-y$ in the integral yields either $A$-independent or $\frac{1}{\Lambda^2}$-terms or terms that vanish exponentially as $\Lambda\to\infty$. Thus, the remaining quadratically divergent \ct $\Lambda^2 f_m^{(1)}$ only leads to terms finite or to be included in the cosmological constant. None of these terms is of any interest here. This glasses diagram thus gives no contribution to the pertinent divergences of the partition function \eqref{eq:z_divergences}. Note that diagrams with a single propagator joining two or three loops were already discarded from the diagrams contributing to the leading divergence in the previous section. \\

The setting sun diagram $~\im{0.25}{new_ct_sunset.pdf}~$ gets two contributions according to which line of the cubic \ct vertex is connected to the bold part of the cubic Liouville vertex. Thus one obtains
\begin{align}
I_{\hbox{\includegraphics[scale=0.10]{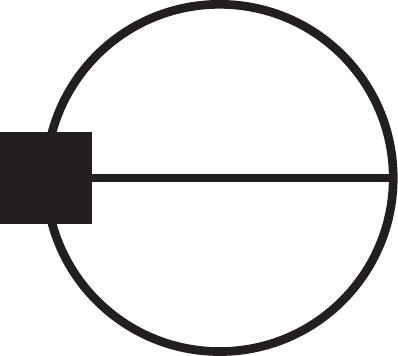}}}~=~\frac{\left(8\pi\right)^2}{\kappa^4}\int&\dxy{x}{y}\nonumber\\
&\left\lbrace f_\phi\widehat{\widetilde{K}}(t_1,x,y)\left[\frac{1}{2}\widehat{\widetilde{K}}\left(t_2,x,y\right)\left(-\frac{\mathrm{d}\widetilde{K}\left(t,x,y\right)}{\mathrm{d}t}\right)_{t=t_3}+\widetilde{K}\left(t_2,x,y\right)\widetilde{K}\left(t_3,x,y\right)\right]\right.\nonumber\\
&+\frac{1}{2}\left[\left(f_\phi+3f_R\right)R_*+3f_m\right]\widehat{\widetilde{K}}\left(t_1,x,y\right)\widehat{\widetilde{K}}\left(t_2,x,y\right)\widetilde{K}\left(t_3,x,y\right)\nonumber\\
&+\left.\frac{R_*}{2}\left(f_RR_*+f_m\right)\widehat{\widetilde{K}}\left(t_1,x,y\right)\widehat{\widetilde{K}}\left(t_2,x,y\right)\widehat{\widetilde{K}}\left(t_3,x,y\right)\right\rbrace \ .
\label{eq:sunset_ct_diagram}
\end{align} 
This leads to the following divergences:
\begin{align}
I_{\hbox{\includegraphics[scale=0.1]{gsoleilnewct.pdf}}}~=~&\frac{1}{\kappa^4}\Bigg\lbrace 6~f_m^{(1)} A\Lambda^2\left(\ln A\Lambda^2\right)^2 + \left[\frac{8}{\alpha_2+\alpha_3}f_\phi^{(1)}+12~f_m^{(1)}\left(b_1-\ln(\alpha_2+\alpha_3)\right)\right] A\Lambda^2\ln A\Lambda^2 \nonumber\\
&+\left[6~f_m^{(2)}+\left(4~f_\phi^{(1)}+6~f_R^{(1)}-6~\alpha_3~f_m^{(1)}\right)AR_*\right]\left(\ln A\Lambda^2\right)^2 \nonumber\\
&+\Bigg[16\Big(\frac{f_\phi^{(1)}}{3}-\alpha_3~f_m^{(1)}\Big)AR_*-\frac{\alpha_2\alpha_3}{\alpha_2+\alpha_3}\left(\frac{8}{\alpha_2+\alpha_3}f_\phi^{(1)}+12~f_m^{(1)}\right)AR_*-48\pi\left(f_\phi^{(1)}-2~\alpha_3~f_m^{(1)}\right) \nonumber\\
&\hskip6.mm+2\left(6~f_m^{(2)}+\left(4~f_\phi^{(1)}+6~f_R^{(1)}-6~\alpha_3~f_m^{(1)}\right)AR_*\right)\left(b_1-\ln(\alpha_2+\alpha_3)\right)\Bigg]\ln A\Lambda^2\Bigg\rbrace
\label{eq:cubic_ct_div}
\end{align} 
where
\begin{align}
b_1=\frac{4\pi}{A_0}\int\mathrm{d}^2x\sqrt{g_0(x)}~\widetilde{G}_\zeta^{A_0}(x)-\gamma-\ln A_0\mu^2
\label{eq:d1}
\end{align}
is a constant independent of $A$. The expression \eqref{eq:cubic_ct_div} is the full contribution from the cubic counterterms to the diverging part of the partition function.

\subsection{Quartic counterterms}
\label{sub:quarticCT}

The quartic \ct action is
\begin{align}
S_{\rm ct}^q=\frac{\left(8\pi\right)^2}{\kappa^4}\frac{1}{2}\int\dx \left[q_\phi\tilde{\phi}^3(\Delta_*-R_*)\tilde{\phi}+\widehat{q}_\phi\tilde{\phi}^2(\Delta_*-2R_*)\tilde{\phi}^2+q_RR_*\tilde{\phi}^4+q_m\tilde{\phi}^4\right]
\label{eq:ct_quartic_action}
\end{align}
with
\begin{align}
q_\phi&=q^{(1)}_\phi~, \nonumber\\
\widehat{q}_\phi&=\widehat{q}^{(1)}_\phi~, \nonumber\\
q_R&=q^{(1)}_R~, \nonumber\\
q_m&=q^{(1)}_m\Lambda^2+\frac{q_m^{(2)}}{A}~.
\label{eq:ct_quartic_values}
\end{align}
Again, the coefficients $q_i^{(j)}$ may depend on the $\alpha_k$ but not on the cutoff $\Lambda$.
This action gives a ``figure-eight'' diagram $~\im{0.25}{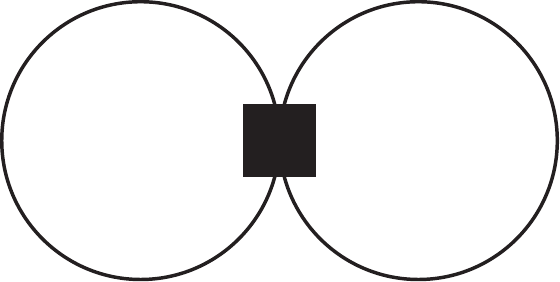}~$:
\begin{align}
I_{\hbox{\includegraphics[scale=0.09]{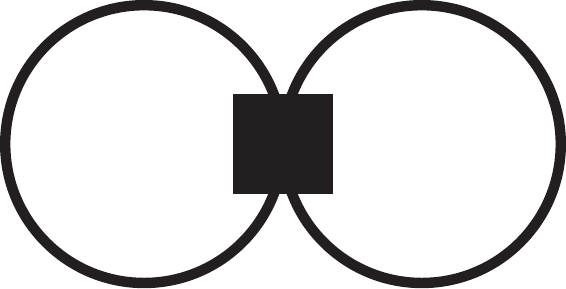}}}~=~ \frac{\left(8\pi\right)^2}{\kappa^4}\int\dx &\left\lbrace -\frac{3}{2}\left(q_\phi\widehat{\widetilde{K}}(t_1,x,x)\widetilde{K}\left(t_2,x,x\right)+\left(q_RR_*+q_m\right)\widehat{\widetilde{K}}\left(t_1,x,x\right)\widehat{\widetilde{K}}\left(t_2,x,x\right)\right)\right.\nonumber\\
&+\left.\widehat{q}_\phi\left(-2\widehat{\widetilde{K}}\left(t_1,x,x\right)\widetilde{K}\left(t_2,x,x\right)+R_*\widehat{\widetilde{K}}\left(t_1,x,x\right)\widehat{\widetilde{K}}\left(t_2,x,x\right)\right)\right\rbrace~,
\label{eq:quartic_ct_diagram}
\end{align}
which contributes as
\begin{align}
I_{\hbox{\includegraphics[scale=0.09]{ghuitnewct.pdf}}}~=~&\frac{1}{\kappa^4}\Bigg\lbrace -6~q_m^{(1)} A\Lambda^2\left(\ln A\Lambda^2\right)^2 - \left[\frac{1}{\alpha_1}\left(6~q_\phi^{(1)}+8~\widehat{q}^{(1)}_\phi\right)+12~q_m^{(1)}\left(b_1-\ln\alpha_1\right)\right] A\Lambda^2\ln A\Lambda^2 \nonumber\\
&-\left[6~q_m^{(2)}+\left(-4~\widehat{q}^{(1)}_\phi+6~q_R^{(1)}\right)AR_*\right]\left(\ln A\Lambda^2\right)^2 
-\Bigg[\Big(\frac{7}{6}AR_*-4\pi\Big)\left(6~q_\phi^{(1)}+8~\widehat{q}_\phi^{(1)}-12~\alpha_1~q_m^{(1)}\right) \nonumber\\
&\hskip10.mm+2\left(6~q_m^{(2)}+\left(-4~\widehat{q}^{(1)}_\phi+6~q_R^{(1)}\right)AR_*\right)\left(b_1-\ln\alpha_1\right)\Bigg]\ln A\Lambda^2\Bigg\rbrace
\label{eq:quartic_ct_div}
\end{align} 
with $b_1$ given in \eqref{eq:d1}.

\subsection{Quadratic two-loop counterterms}
\label{sub:quadratcCT}

The quadratic counterterms \eqref{eq:ct_action} did contribute via one-loop diagrams to the two-loop partition function, but also via two-loop diagrams to the three-loop partition function as shown in the above computation. However, as always, the counterterm coefficients get contributions at different orders in perturbation theory. If we call $c_\phi$, $c_R$ and $c_m$ the coefficients in \eqref{eq:old_ct_values}, we may add to them an additional piece $\frac{1}{\kappa^2} c'_\phi$, $\frac{1}{\kappa^2} c'_R$ and $\frac{1}{\kappa^2} c'_m$, so that $c_\phi^{\text{tot}}=c_\phi+\frac{1}{\kappa^2}c'_\phi+\mathcal{O}(\frac{1}{\kappa^4})$, etc. Overall, the $c'$ are accompanied by a factor $\frac{1}{\kappa^4}$ and they contribute via one-loop diagrams to the three-loop partition function. Thus we also add the following counterterm action
\begin{align}
S^{\text{quad'}}_{\rm ct}=\frac{8\pi}{\kappa^4}\int \dx\left[\frac{c'_\phi}{2}\tilde{\phi}(\Delta_*-R_*)\tilde{\phi}+\frac{c'_R}{2}R_*\tilde{\phi}^2+\frac{c'_m}{2}\tilde{\phi}^2\right] \ ,
\label{eq:ct_action}
\end{align}
where, again,
\begin{align}
c'_\phi&=c'^{(1)}_\phi ~,\nonumber\\
c'_R&=c'^{(1)}_R ~,\nonumber\\
c'_m&=c'^{(1)}_m\Lambda^2 +\frac{c'^{(2)}_m}{A} ~.
\label{eq:old_ct_k4}
\end{align}
The \ct action \eqref{eq:ct_action} then provides a new one-loop diagram of order $\kappa^{-4}$: $~\im{0.35}{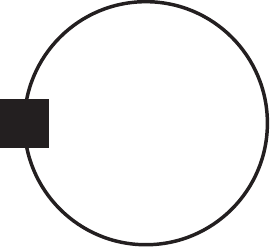}$
\begin{align}
I_{\hbox{\includegraphics[scale=0.12]{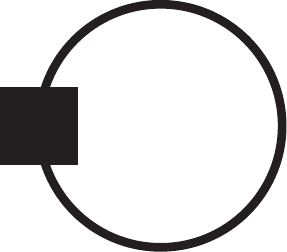}}}~=~-\frac{1}{2}\frac{\left(8\pi\right)}{\kappa^4}\int\dx\left[c'_\phi \widetilde{K}(t,x,x)+\left(c'_RR_*+c'_m\right)\widehat{\widetilde{K}}(t,x,x)\right]
\label{eq:old_new_ct_diagram}
\end{align} 
leading to the following divergences:
\begin{align}
I_{\hbox{\includegraphics[scale=0.12]{goldct.pdf}}}~=~\frac{1}{\kappa^4}\left(-c'^{(1)}_m A\Lambda^2\ln A\Lambda^2-\left(c'^{(2)}_m+c'^{(1)}_RR_*A\right)\ln A\Lambda^2\right)~.
\label{eq:old_new_ct_div}
\end{align}

Moreover, two parameters of the two-loop counterterms  \eqref{eq:old_ct_values} are still unconstrained: $\widehat{c}_\phi$ and $\widehat{c}_R$. Although only $\widehat{c}_R$ appears in the two-loop partition function, both may contribute to the divergent part of the partition function at three loops, through the diagrams $~\im{0.18}{gmct.pdf}~$, $~\im{0.18}{gctct.pdf}~$, $~\im{0.17}{ghuitct.pdf}~$ and $~\im{0.15}{gsoleilct.pdf}~$. Their diverging contributions are displayed below:
\begin{align}
I_{\hbox{\includegraphics[scale=0.1]{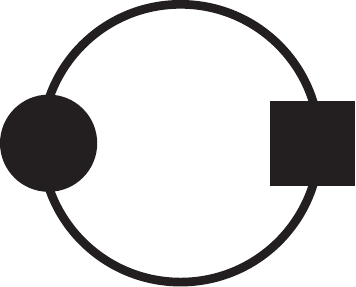}}}~=~&\frac{1}{\kappa^4}\Bigg\lbrace -\frac{2}{\alpha_1}~\widehat{c}_\phi~ A\Lambda^2\ln A\Lambda^2 - 2~\Big(\frac{7}{6}AR_*-4\pi\Big)~\widehat{c}_\phi~\ln A\Lambda^2\Bigg\rbrace~, 
\nonumber\\
I_{\hbox{\includegraphics[scale=0.1]{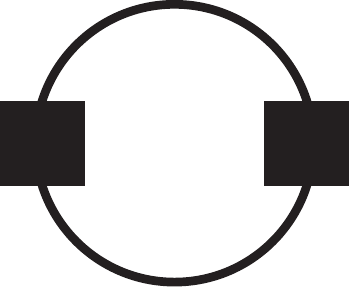}}}~=~&\frac{1}{\kappa^4}\Bigg\lbrace \left(-\frac{10}{\alpha_1}+\frac{8}{\alpha_1+\alpha_2}\right)~\widehat{c}_\phi~ A\Lambda^2\ln A\Lambda^2 +\Bigg[~2AR_*~\widehat{c}_\phi\frac{\widehat{c}_R}{\pi}+2\left(\frac{7}{6}AR_*-4\pi\right)\frac{\widehat{c}_R}{2\pi} 
\nonumber\\
&\hskip10.mm+4AR_*\left(\widehat{c}_\phi+\frac{\widehat{c}_R}{2\pi}\right)\Bigg(3~\Big(\ln(\alpha_1+\alpha_2)-\ln\alpha_1\Big)-\frac{19}{12}-\frac{2\alpha_1\alpha_2}{(\alpha_1+\alpha_2)^2}\Bigg)\Bigg]\ln A\Lambda^2\Bigg\rbrace~, 
\nonumber\\
I_{\hbox{\includegraphics[scale=0.1]{ghuitct.pdf}}}~=~&\frac{1}{\kappa^4}\Bigg\lbrace 12\left(\frac{1}{\alpha_1}+\frac{1}{\alpha_1+\alpha_2}\right)~\widehat{c}_\phi~ A\Lambda^2\ln A\Lambda^2 +12AR_*\left(\widehat{c}_\phi+\frac{\widehat{c}_R}{2\pi}\right)\left(\ln A\Lambda^2\right)^2 
\nonumber\\
&\hskip10.mm+\Bigg[~12AR_*\left(\widehat{c}_\phi+\frac{\widehat{c}_R}{2\pi}\right)\Big(2~b_1-\ln\alpha_1-\ln(\alpha_1+\alpha_2)\Big)+24~\Big(\frac{7}{6}AR_*-4\pi\Big)\widehat{c}_\phi
\nonumber\\
&\hskip10.mm+24AR_*^2~\widehat{c}_R\int\mathrm{d}^2x\mathrm{d}^2y\sqrt{g_*(x)g_*(y)}~\widehat{\widetilde{K}}\left(t_1,x,y\right)\widehat{\widetilde{K}}\left(t_2,x,y\right)\Bigg]\ln A\Lambda^2\Bigg\rbrace~,
\label{eq:ct2loopsdiv_1}
\end{align}
and
\begin{align}
I_{\hbox{\includegraphics[scale=0.08]{gsoleilct.pdf}}}~=~&\frac{1}{\kappa^4}\Bigg\lbrace - 8 \Big(\frac{1}{\alpha_1+\alpha_2}+\frac{2}{\alpha_1+\alpha_2+\alpha_3}\Big)~\widehat{c}_\phi~ A\Lambda^2\ln A\Lambda^2 -\Big(18~\widehat{c}_\phi+7~\frac{\widehat{c}_R}{\pi}\Big)\left(\ln A\Lambda^2\right)^2
\nonumber\\
&\hskip5.mm+\Bigg[-24AR_*^2~\widehat{c}_R\int\mathrm{d}^2x\mathrm{d}^2y\sqrt{g_*(x)g_*(y)}~\widehat{\widetilde{K}}\left(t_1,x,y\right)\widehat{\widetilde{K}}\left(t_2,x,y\right)
-2\Big(18~\widehat{c}_\phi+7~\frac{\widehat{c}_R}{\pi}\Big)b_1
\nonumber\\
&\hskip12.mm +4AR_*\Big[\Big(\widehat{c}_\phi+\frac{\widehat{c}_R}{2\pi}\Big)\Big(\frac{2\alpha_1\alpha_2}{(\alpha_1+\alpha_2)^2}+3\ln(\alpha_1+\alpha_2)\Big)
+6\Big(\widehat{c}_\phi+\frac{\widehat{c}_R}{3\pi}\Big)\ln(\alpha_1+\alpha_2+\alpha_3)\Big]\nonumber\\
&\hskip12.mm +4\Big[36\pi~\widehat{c}_\phi+AR_*\Big(\frac{\widehat{c}_R}{2\pi}-\Big(\frac{8}{3}+\frac{14}{9}\frac{\alpha_1^2+\alpha_2^2+\alpha_3^2}{(\alpha_1+\alpha_2+\alpha_3)^2}\Big)\widehat{c}_\phi\Big)\Big] \Bigg]\ln A\Lambda^2\Bigg\rbrace~.
\label{eq:ct2loopsdiv_2}
\end{align}
None of these contains a $A\Lambda^2 (\ln A\Lambda^2)^2$ divergence and this is why these finite counterterm coefficients $\widehat c_\phi$ and $\widehat c_R$ did not contribute to our computation in section~3.

\subsection{Total counterterm contribution to the partition function}
\label{sub:totalpartfctCT}

Since the glasses diagram has no divergence other than in $A\Lambda^2$, the total contribution one could get from the counterterms  to the three-loop partition function is given by summing \eqref{eq:cubic_ct_div}, \eqref{eq:quartic_ct_div}, \eqref{eq:old_new_ct_div}, \eqref{eq:ct2loopsdiv_1} and \eqref{eq:ct2loopsdiv_2}. Recalling $A R_*=8\pi(1-h)$, cf.~\eqref{eq:scaling_relations}, one gets:
\begin{align}
\ln Z[A]^{\text{CT}}_{3-\text{loop}}
=&~\frac{1}{\kappa^4} \Bigg\lbrace \Omega_1~A\Lambda^2\left(\ln A\Lambda^2\right)^2+\Omega_2~A\Lambda^2\ln A\Lambda^2+\Omega_3~ 
\left(\ln A\Lambda^2\right)^2+\Omega_4~\ln A\Lambda^2 \Bigg\rbrace
\label{eq:z_ct_div}
\end{align}
with
\begin{align}
&\Omega_1 =~ 6\left(f_m^{(1)}-q_m^{(1)}\right)
\nonumber\\
&\Omega_2 =-\frac{16}{\alpha_1+\alpha_2+\alpha_3}~\widehat{c}_\phi+~\frac{1}{\alpha_1+\alpha_2}\left(8 ~f_\phi^{(1)}+12~\widehat{c}_\phi\right) - \frac{1}{\alpha_1}\left(6~q_\phi^{(1)}+8~\widehat{q}^{(1)}_\phi\right) - c'^{(1)}_m
\nonumber \\
&\hskip9.mm +2~\Omega_1~b_1
- 12\left(f_m^{(1)}\ln(\alpha_1+\alpha_2)-q_m^{(1)}\ln\alpha_1\right) 
\nonumber\\
&\Omega_3 =~\Omega_3^{(a)}+\Omega_3^{(b)}+\Omega_3^{(c)}
\nonumber\\
&\Omega_3^{(a)} =-6~q_m^{(2)}+\left(4~\widehat{q}^{(1)}_\phi+12\left(\widehat{c}_\phi+\frac{\widehat{c}_R}{2\pi}\right)-6~q_R^{(1)}\right)8\pi\left(1-h\right)
\nonumber\\
&\Omega_3^{(b)}(\alpha_1) =~6~f_m^{(2)}+\left(4~f_\phi^{(1)}-6\left(\widehat{c}_\phi+\frac{\widehat{c}_R}{2\pi}-f_R^{(1)}+\alpha_1~f_m^{(1)}\right)\right) 8\pi\left(1-h\right)
\nonumber\\
&\Omega_3^{(c)} =-12\left(\widehat{c}_\phi+\frac{\widehat{c}_R}{3\pi}\right)~8\pi\left(1-h\right)
\nonumber\\
&\Omega_4 =~ 2~\Omega_3~b_1- 2\left(~\Omega_3^{(a)}\ln\alpha_1+\Omega_3^{(b)}(\alpha_1)\ln(\alpha_1+\alpha_2)+\Omega_3^{(c)}\ln(\alpha_1+\alpha_2+\alpha_3) \right)
\nonumber\\
&\hskip9.mm+4\pi\left(1-\frac{7}{3}\left(1-h\right)\right)\Big(6~q_\phi^{(1)}+8~\widehat{q}_\phi^{(1)}+12~\alpha_1~f_m^{(1)}+12~\widehat{c}_\phi-12~f_\phi^{(1)}+2~\alpha_1~\Omega_1\Big)
\nonumber\\
&\hskip9.mm+8\pi\left(1-h\right)\Bigg(\frac{26}{3}\left(\widehat{c}_\phi-f_\phi^{(1)}\right)+\frac{12\,\alpha_1^2}{\alpha_1+\alpha_2}f_m^{(1)}-\frac{8\,\alpha_1\alpha_2}{\left(\alpha_1+\alpha_2\right)^2}f_\phi^{(1)}+\frac{56}{3}\frac{\alpha_1^2}{(\alpha_1+\alpha_2+\alpha_3)^2}\widehat{c}_\phi\Bigg)
\nonumber\\
&\hskip9.mm+8\pi\left(1+14\left(1-h\right)\right)\widehat{c}_\phi-\left( c'^{(2)}_m +4~\widehat{c}_R + 8\pi\left(1-h\right)c'^{(1)}_R \right)+16\left(1-h\right)\,\widehat{c}_\phi\,\widehat{c}_R
\label{eq:z_ct_coeff}
\end{align}
where $b_1$ was defined in \eqref{eq:d1}.

This is the total contribution  to the three-loop partition function of the  counterterms that have not been previously fixed by the order $\frac{1}{\kappa^2}$ (``two-loop'') computation of \cite{Bilal:2014mla}. Requiring the one-loop two-point function to be finite and regulator independent  fixed $c_m$ and parts of $c_\phi$ and $c_R$. Thus, only their so-far undetermined regularization-independent parts $\widehat{c}_\phi$ and $\widehat{c}_R$ have been included in \eqref{eq:z_ct_coeff}. 

One way to determine some of these \cts is to compute the two-loop two-point function (order $\frac{1}{\kappa^4}$) and the one-loop three-point function (order $\frac{1}{\kappa^3}$) and one-loop four-point function (order $\frac{1}{\kappa^4}$)  and to require them to be finite and regularization independent. Imposing finiteness will completely determine certain combinations of the counterterm coefficients, while imposing regularization independence of the finite terms will fix certain other combinations up to constants.
 
The computations of the two-loop two-point function and of the one-loop three-point and four-point functions clearly are beyond the scope of this work. However, without actually doing this computation, there are still interesting remarks we can make. We can rather easily determine the contributions of the counterterms to these $n$-point functions. This will  tell us which combinations of the counterterm coefficient would be fixed by such a computation. We will  find that the relevant combinations are indeed the same as those appearing in the $\Omega_i$ of the three-loop partition function. Although ``expected'', this is by no means obvious and constitutes a nice consistency check. 

It is straightforward to see that the cubic and quartic counterterms contribute to the diverging parts of the three- and four-point functions as
\begin{align}
\left.\vcenter{\hbox{\includegraphics[scale=0.13]{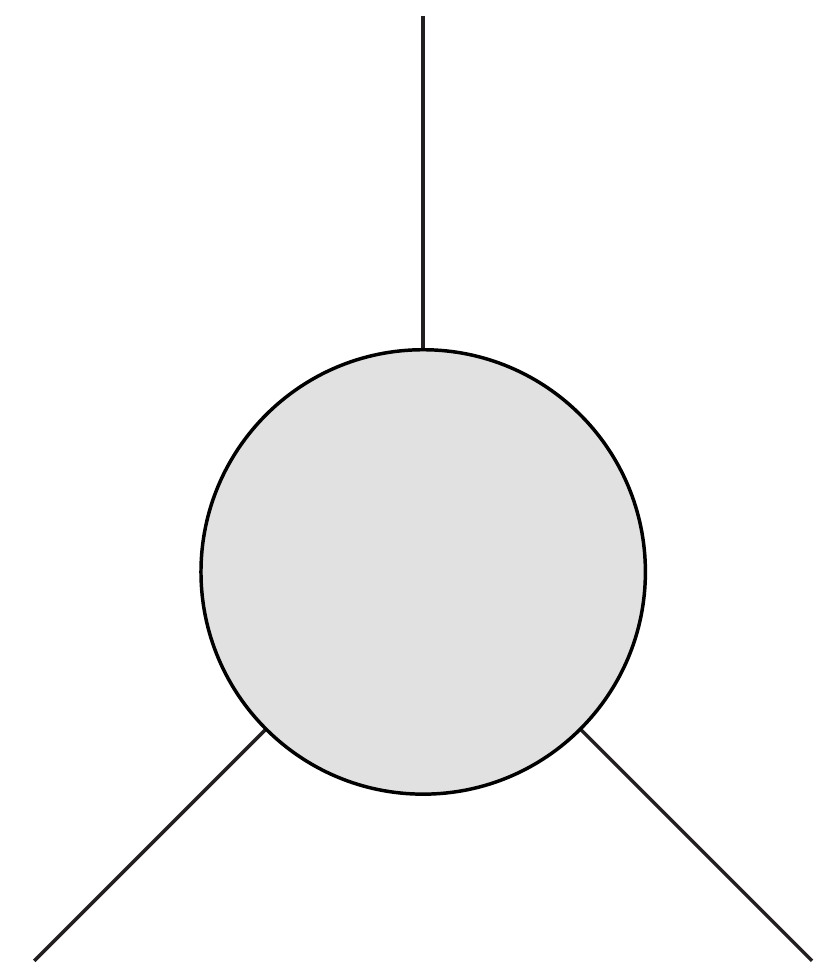}}}\right|_{\text{div}}^{\text{CT}}&=-\frac{48\pi^{3/2}}{\kappa^2}f_m^{(1)}\Lambda^2\int\mathrm{d}^2x\sqrt{g_*}~\widehat{\widetilde{K}}\left(t_1,a,x\right)\widehat{\widetilde{K}}\left(t_2,b,x\right)\widehat{\widetilde{K}}\left(t_3,c,x\right)~, 
\nonumber\\ 
\left.\vcenter{\hbox{\includegraphics[scale=0.13]{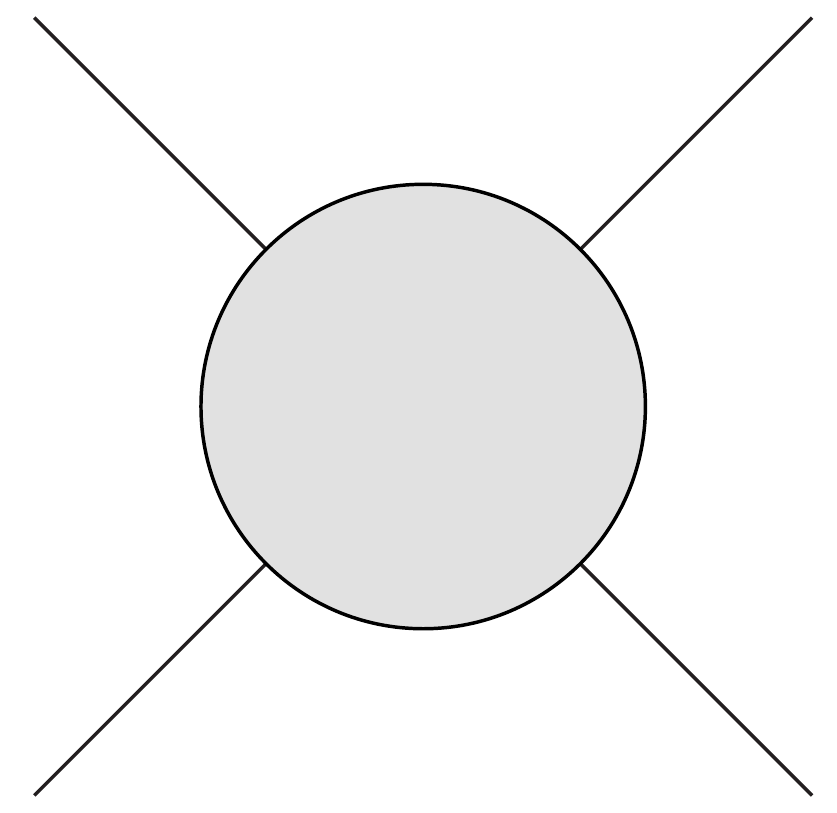}}}\right|_{\text{div}}^{\text{CT}}&=- 12\frac{(8\pi)^2}{\kappa^4} 
q_m^{(1)}\Lambda^2\int\mathrm{d}^2x\sqrt{g_*}~\widehat{\widetilde{K}}\left(t_1,a,x\right)\widehat{\widetilde{K}}\left(t_2,b,x\right)\widehat{\widetilde{K}}\left(t_3,c,x\right)\widehat{\widetilde{K}}\left(t_4,d,x\right)~.
\label{fcts3et4pts}
\end{align}
Thus finiteness of these functions fixes both $f_m^{(1)}$ and $q_m^{(1)}$ and hence, $\Omega_1$. Finiteness of the two-point function at one loop (order $\frac{1}{\kappa^2}$) was already imposed in \cite{Bilal:2014mla} and resulted in the determination of $c_m$ to this order. Here we will only consider its two-loop $\frac{1}{\kappa^4}$  part. We find that the contributions of the counterterms to the diverging part of the two-loop two-point function is
\begin{align}
\left.\vcenter{\hbox{\includegraphics[scale=0.13]{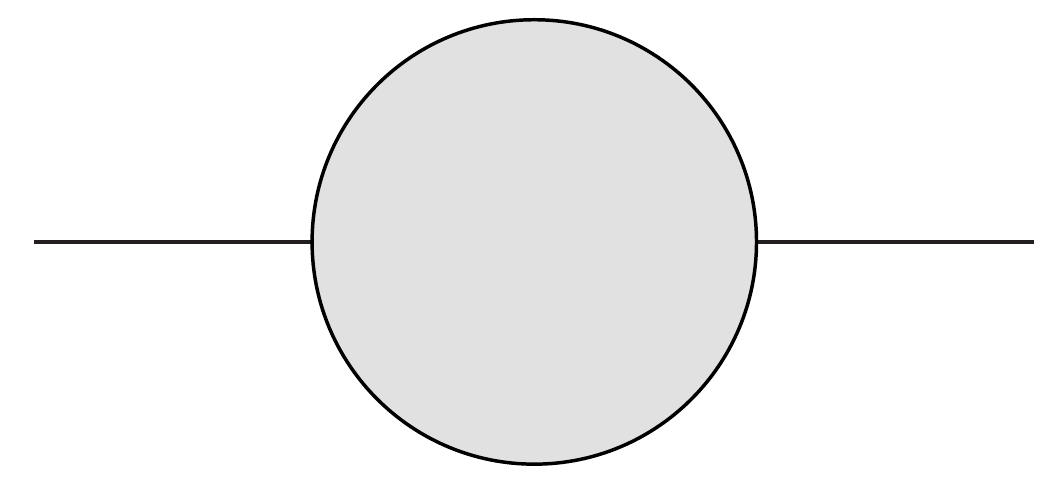}}}\right|_{\frac{1}{\kappa^4}, \text{div}}^{\text{CT}}=~\frac{8\pi}{\kappa^4}\Bigg\lbrace 
&\Big(\rho_1~\Lambda^2\ln A\Lambda^2+\rho_2~\Lambda^2+\rho_3\,\frac{\ln A\Lambda^2}{A}\Big)\int\mathrm{d}^2x\sqrt{g_*}~\widehat{\widetilde{K}}\left(t_1,a,x\right)\widehat{\widetilde{K}}\left(t_2,b,x\right)
\nonumber\\
&+\rho_4~\Lambda^2\int\mathrm{d}^2x\sqrt{g_*}~\tilde{G}_\zeta^{A_0}(x)\widehat{\widetilde{K}}\left(t_1,a,x\right)\widehat{\widetilde{K}}\left(t_2,b,x\right)
\nonumber\\
&+\rho_5\,\ln A\Lambda^2~\widehat{\widetilde{K}}\left(t_1,a,b\right) 
\Bigg\rbrace~,
\label{fct2pts}
\end{align}
with
\begin{align}
\rho_1=2~\Omega_1 \; , \quad \rho_2&=\Omega_2 + F\left[\alpha_i,f^{(1)}_m,q^{(1)}_m\right] \; , \quad \rho_3=2\left(\Omega_3+6~\alpha_2~f_m^{(1)}AR_*\right) \; , \nonumber\\
\rho_4=8\pi~\Omega_1 \; , \quad &\rho_5=-\left(6~q_\phi^{(1)}+8~\widehat{q}_\phi^{(1)}+12~\alpha_1~f_m^{(1)}+12~\widehat{c}_\phi-12~f_\phi^{(1)}\right)~.
\label{rhoOmega}
\end{align} 
The expression $F\left[\alpha_i,f^{(1)}_m,q^{(1)}_m\right]$ can be computed straightforwardly, but its exact form is irrelevant for the present discussion. 
Finiteness of the two-point function at order $\frac{1}{\kappa^4}$ then fixes all combinations $\rho_1, \ldots \rho_5$. Since $F$ is a known function of the already determined $f_m^{(1)}$and $q_m^{(1)}$ this then fixes  $\Omega_2$ and $\Omega_3$, as well as the combination $6~q_\phi^{(1)}+8~\widehat{q}_\phi^{(1)}+12~\widehat{c}_\phi-12~f_\phi^{(1)}$.

\vskip2.mm

Thus, all the coefficients $\Omega_1$, $\Omega_2$ and $\Omega_3$ of the diverging parts of the counterterm contributions to the partition function \eqref{eq:z_ct_div} are exactly determined by the requirement of the finiteness of the two-loop two-point function and of the one-loop three-point and four-point functions~! Obviously, we expect this determination to be such that \eqref{eq:z_ct_div} precisely cancels the divergences of the genuine three-loop part of this partition function, as was indeed the case for the two-loop computation of \cite{Bilal:2014mla}.

Let us next discuss $\Omega_4$ which is the counterterm contribution to the order $\frac{1}{\kappa^4}$ part of $\gamma_{\rm str}$. With the $f^{(1)}_m$, $q^{(1)}_m$ and the $\rho_i$ been fixed, also $\Omega_1$, $\Omega_2$ and $\Omega_3$ are fixed and we consider $\Omega_3^{(a)}$ as a function of 
$\Omega_3^{(b)}$ and $\Omega_3^{(c)}$, i.e. of $\Omega_3^{(b)}$, $\widehat{c}_\phi$ and $\widehat{c}_R$. (Note that the second line in the expression of $\Omega_4$ can be expressed though $\rho_5$ and $\Omega_1$.) Thus $\Omega_4$ depends on the following six undetermined constants:  $\Omega_3^{(b)}$,  $\widehat{c}_\phi$, $\widehat{c}_R$, $f^{(1)}_\phi$, $c'^{(2)}_m$ and $c'^{(1)}_R$.   

Furthermore, one may require  the ``strong locality condition'' that the non-local terms in the measure \eqref{eq:measure_action_3} and \ct actions \eqref{eq:ct_action}, \eqref{eq:ct_cubic_action}, \eqref{eq:ct_quartic_action} cancel out. This  fixes $q^{(2)}_m$, $f^{(2)}_m$ and $c'^{(2)}_m$ as
\begin{align}
q^{(2)}_m = -1~,~~~~~~ f_m^{(2)} = -\frac{4}{3} ~,~~~~~~ c'^{(2)}_m = 0~,
\label{eq:locality_ct}
\end{align}
since the corresponding $\frac{1}{A}$ terms in \eqref{eq:old_ct_values}, \eqref{eq:ct_cubic_values} and \eqref{eq:ct_quartic_values} are 
\begin{align}
\frac{4\pi}{A}\left[\frac{2\sqrt{\pi}}{\kappa^3}\left(f_m^{(2)}+\frac{4}{3}\right)\tilde{\phi}^3-\frac{1}{\kappa^4}c'^{(2)}_m\tilde{\phi}^2+\frac{8\pi^2}{\kappa^4}\left(q^{(2)}_m+1\right)\tilde{\phi}^4\right]=0~.
\label{eq:locality_arg}
\end{align}
Thus, among the six undetermined constants in $\Omega_4$, only $c'^{(2)}_m$ is fixed, and we end up with five free finite renormalization constants on which $\Omega_4$ depends: $\Omega_3^{(b)}$, $\widehat{c}_\phi$, $\widehat{c}_R$, $f^{(1)}_\phi$, and $c'^{(1)}_R$. We conclude that in addition to the undetermined $\widehat{c}_R$ which already entered as a free parameter in the two-loop expression of $\gamma_{\rm str}$, at three loops $\gamma_{\rm str}$ depends on four additional undetermined parameters.

\vskip2.mm

Finally, as anticipated in section 3, none of the counterterms  contributes to the $\left(\ln A\Lambda^2\right)^3$ divergence. The only way to generate such divergences would be by introducing non-local counterterm coefficients that already involve a factor of  $\ln A\Lambda^2$. However, as repeatedly argued, such counterterms should be forbidden. Then, since there is no possible counterterm for a $\left(\ln A\Lambda^2\right)^3$ divergence,
such a divergence is  required to cancel in the first place between the three-loop vacuum diagrams. As shown above, this is indeed the case.

\section{Discussion}
\label{sec:discussion}

The purpose of our work was to check if and which new counterterms  are required  at three loops. We have therefore computed the leading divergence of the three-loop partition function at fixed area, cf \eqref{eq:result_leading}. It does not vanish  and thus genuine three-loop counterterms  are required. It is interesting to note that the two-loop computation already pointed to the insertion of new counterterms  at three loops. Indeed, the counterterms  inserted at two loops have a strong similarity with the measure terms at two loops. Yet, at three loops, the measure action gives rise to cubic and quartic vertices unlike the two-loop \ct vertices. Therefore, one could have expected additional counterterms  to be needed. This argument can be generalized to all orders, as the measure action gets additional structures at every order in the loop expansion. If the counterterms  are to be understood as a renormalization of the measure action, the latter itself coming from the regularization of the measure for the metrics, then new counterterms  have to be introduced at every order in the perturbation series. On the other hand, what is really surprising and encouraging is that if one requires the counterterms  to be local, in particular that no \ct coefficient with a $\ln A\Lambda^2$ divergence is allowed, then all the divergences may be offset but the $\left(\ln A\Lambda^2\right)^3$ divergence. However, as we showed, this divergence cancels out between the three-loop diagrams, meaning that local counterterms  are enough to balance all the non-local divergences. Moreover, the required \ct action has a structure similar to those of the measure action, supporting the understanding of counterterms  as a renormalization of the measure.

Nevertheless, with no other way to discriminate the counterterms  than to forbid $(\ln A\Lambda^2)$-like non-local terms, many new free parameters appear. At three loops, doing so gives rise to twelve new parameters. Imposing the divergences to vanish in the one-loop three- and four-point functions and in the two-loop two-point function fixes two  parameters and three combinations of the parameters. We found that with these parameters and combinations of parameters fixed, the diverging part of the three-loop partition function is also completely fixed with no additional adjustable parameter remaining. Obviously, as was the case at two loops, we expect this to happen in precisely such a way that all divergences in the three-loop partition function  cancel, except for the $(\ln A\Lambda^2)$-piece that yields the three-loop contribution to the string susceptibility. Indeed, this is the only coefficient of the three-loop partition function which contains undetermined finite renormalization constants. More precisely, it depends on six  unconstrained renormalization constants. We argued that there are two different notions of locality of the counterterm coefficients: while coefficients involving $\ln A\Lambda^2$ were excluded, we did allow coefficients proportional to  $\frac{1}{A}$ since such non-local terms already appeared through the measure action. Introducing such $\frac{1}{A}$ counterterms in precisely such a way as to cancel the corresponding $\frac{1}{A}$ terms in the measure action was referred to as   ``strong locality condition''. Imposing this condition fixes one of the six free parameters in the contribution to the string susceptibility, leaving us with five free renormalization constants on which the three-loop contribution to $\gamma_{\rm str}$ depends. One of these free renormalization constants was already present in the two-loop string susceptibility, so that at three-loops, four new constants play a role.

Several additional requirements should be considered, such as the condition that neither the $n$-point functions nor the partition function should depend on the choice of regularization. In particular, the regularization function $\varphi(\alpha_i)$ satisfies $\int_0^\infty\mathrm{d}\alpha_i\varphi(\alpha_i)=1$ and certain regularity conditions at $0$ and infinity, but is otherwise arbitrary.  Its choice should not impact any final, physical result. This  means that all the dependence in the $\alpha_i$ must disappear in the end. Although important, this argument is not enough to fully determine the counterterms, in particular it cannot fix any $\alpha$-independent pieces.  Another criterion is the background independence. Physical results must not depend on  the background metric $g_0$ arbitrarily chosen to write the Liouville action, define the conformal factor $\sigma$ and thus the Kähler field $\tilde{\phi}$. The eigenmodes of $\Delta_*$ are also defined through this choice of the reference metric. One way to check for background independence is to derive the cocycle identities for the various actions. It is easy to check that the Liouville action satisfies this condition. Formally, the same is true for the measure action. However, as usual, the need to introduce an explicit regularization, making reference to a background metric, obscures the background independence and makes it difficult to be verified. It could well be that some indirect criterion for background independence fixes some or all of our free renomalization constants. 

\vskip5.mm
\noindent
{\Large \bf Acknowledgements}

\vskip10.mm
\noindent
L.L. is grateful to the Capital Fund Management foundation for a Ph.D. fellowship.

\newpage

\begin{appendix}

\section{Appendix}
\label{sec:appendix}

When computing the diagrams $\im{0.22}{gbonneige.pdf}~$, $\im{0.18}{ggoggles.pdf}~$, $\im{0.2}{gcylindre.pdf}$ and$~\im{0.16}{gtetra.pdf}~$, one gets infinite series of terms contributing to the coefficients of $A\Lambda^2\left(\ln A\Lambda^2\right)^2$. The details of the computation of these series are described hereafter. For the sake of brevity, notational short-cuts are defined: $\dnu(x)=\mathrm{d}^2x\sqrt{g_*(x)}$, $\hk_{i}(x,y)=\hk(t_i,x,y)$ and $\hk_{i,j}(x,y)=\hk(t_i+t_j,x,y)$. 

Among the series of diverging contributions appearing in these four diagrams, some may be straightforwardly computed by Taylor expanding the terms, as was done in section \ref{sec:divergence}. These integrals are hereafter noted by $J$. Namely, they are:
\begin{align}
J^{(1)}_{\hbox{\includegraphics[scale=0.09]{ggoggles.pdf}}}=&\int \dnu(x)\dnu(y)\dnu(z)~\hk_{1}(x,z)\kt_{2}(x,z)\hk_{3}(y,z)\kt_{4}(y,z)\kt_{5}(x,y) ~,\nonumber\\
J^{(2)}_{\hbox{\includegraphics[scale=0.09]{ggoggles.pdf}}}=&\int \dnu(x)\dnu(y)\dnu(z)~\hk_1(x,z)\hk_2(x,z)\hk_3(y,z)\kt_4(y,z)\left(-\frac{\mathrm{d}}{\mathrm{d}t_5}\kt_5(x,y)\right) ~,\nonumber\\
J^{(3)}_{\hbox{\includegraphics[scale=0.09]{ggoggles.pdf}}}=&\int \dnu(x)\dnu(y)\dnu(z)~\hk_1(x,z)\left(-\frac{\mathrm{d}}{\mathrm{d}t_2}\kt_2(x,z)\right)\hk_3(y,z)\kt_4(y,z)\hk_5(x,y) ~,\nonumber\\
J^{(4)}_{\hbox{\includegraphics[scale=0.09]{ggoggles.pdf}}}=&\int \dnu(x)\dnu(y)\dnu(z)~\hk_1(x,z)\hk_2(x,z)\hk_3(y,z)\left(-\frac{\mathrm{d}}{\mathrm{d}t_4}\kt_4(y,z)\right)\kt_5(x,y) ~,\nonumber\\
J^{(1)}_{\hbox{\includegraphics[scale=0.12]{gbonneige.pdf}}}=&\int \dnu(x)\dnu(y)\dnu(z)~\hk_1(z,z)\kt_2(x,z)\hk_3(x,y)\kt_4(x,y)\kt_5(y,z) ~,\nonumber\\
J^{(2)}_{\hbox{\includegraphics[scale=0.12]{gbonneige.pdf}}}=&\int \dnu(x)\dnu(y)\dnu(z)~\hk_1(z,z)\kt_2(x,z)\hk_3(x,y)\left(-\frac{\mathrm{d}}{\mathrm{d}t_4}\kt_4(x,y)\right)\hk_5(y,z) ~,\nonumber\\
J^{(3)}_{\hbox{\includegraphics[scale=0.12]{gbonneige.pdf}}}=&\int \dnu(x)\dnu(y)\dnu(z)~\hk_1(z,z)\hk_2(x,z)\hk_3(x,y)\kt_4(x,y)\left(-\frac{\mathrm{d}}{\mathrm{d}t_5}\kt_5(y,z)\right) ~,\nonumber\\
J^{(1)}_{\hbox{\includegraphics[scale=0.1]{gcylindre.pdf}}}=& \int \dnu(x)\dnu(y)\dnu(z)\dnu(w)~\hk_{1}(x,y)\kt_{2}(x,y)\kt_{3}(y,z)\hk_{4}(z,w)\kt_{5}(z,w)\kt_6(x,w)~,\nonumber\\
J^{(2)}_{\hbox{\includegraphics[scale=0.1]{gcylindre.pdf}}}=& \int \dnu(x)\dnu(y)\dnu(z)\dnu(w)~\hk_{1}(x,y)\left(-\frac{\mathrm{d}}{\mathrm{d}t_2}\kt_2(x,y)\right)\hk_{3}(y,z)\hk_{4}(z,w)\left(-\frac{\mathrm{d}}{\mathrm{d}t_5}\kt_5(z,w)\right)\hk_6(x,w) ~,\nonumber\\
J^{(3)}_{\hbox{\includegraphics[scale=0.1]{gcylindre.pdf}}}=& \int \dnu(x)\dnu(y)\dnu(z)\dnu(w)~\hk_{1}(x,y)\kt_2(x,y)\hk_{3}(y,z)\hk_{4}(z,w)\kt_5(z,w)\left(-\frac{\mathrm{d}}{\mathrm{d}t_6}\kt_6(x,w)\right)~,\nonumber\\
J^{(4)}_{\hbox{\includegraphics[scale=0.1]{gcylindre.pdf}}}=& \int \dnu(x)\dnu(y)\dnu(z)\dnu(w)~\hk_{1}(x,y)\kt_2(x,y)\hk_{3}(y,z)\hk_{4}(z,w)\left(-\frac{\mathrm{d}}{\mathrm{d}t_5}\kt_5(z,w)\right)\kt_6(x,w)~.
\label{apx:J_int}
\end{align}
Note that $J^{(1)}_{\hbox{\includegraphics[scale=0.09]{ggoggles.pdf}}}$ was already computed in section \ref{sec:divergence}. It is useful to define:
\begin{align}
&B_n(t_a,t_b,t_c;x)=\int \dnu(z)~\hk_a(x,z)\hk_b(x,z)\frac{\mathrm{d}^n}{\mathrm{d}t_c^n}\kt_c(x,z)~, \nonumber\\
&C_{n,m}(t_a,t_b,t_c;x)=\int \dnu(z)~\hk_a(x,z)\frac{\mathrm{d}^n}{\mathrm{d}t_b^n}\kt_b(x,z)\frac{\mathrm{d}^m}{\mathrm{d}t_c^m}\kt_c(x,z)~, \nonumber\\
&D_n(t_a,t_b,t_c;x)=\int \dnu(z)~\hk_a(z,z)\hk_b(x,z)\frac{\mathrm{d}^n}{\mathrm{d}t_c^n}\kt_c(x,z)~, \nonumber\\
&E_{n,m}(t_a,t_b,t_c;x)=\int \dnu(z)~\hk_a(z,z)\frac{\mathrm{d}^n}{\mathrm{d}t_b^n}\kt_b(x,z)\frac{\mathrm{d}^m}{\mathrm{d}t_c^m}\kt_c(x,z)~.
\label{apx:def_int_ref_xy}
\end{align}
The only divergence investigated here is the one in $A\Lambda^2\left(\ln A\Lambda^2\right)^2$, which cannot appear unless two $\hk$s are present, since the logarithmic divergence comes from such terms \eqref{eq:hk_xx}. The terms without at least two $\hk$s after doing the expansions are thus discarded in the following. Remembering that $t=\frac{\alpha}{\Lambda^2}$, the previous integrals may then be rewritten as:
\begin{align}
J^{(1)}_{\hbox{\includegraphics[scale=0.09]{ggoggles.pdf}}}=&\int \dnu(x)~\hk_{3,4}(x,x)\sum\limits_{n=0}^\infty \frac{t_4^n}{n!}C_{0,n}(t_1,t_2,t_5;x)~,\nonumber\\
J^{(2)}_{\hbox{\includegraphics[scale=0.09]{ggoggles.pdf}}}=&-\int \dnu(x)~\hk_{3,4}(x,x)\sum\limits_{n=0}^\infty \frac{t_4^{n}}{n!}B_{n+1}(t_1,t_2,t_5;x)~,\nonumber\\
J^{(3)}_{\hbox{\includegraphics[scale=0.09]{ggoggles.pdf}}}=&\int \dnu(x)~\hk_{3,4}(x,x)\left[-B_1(t_1,t_2,t_5;x)+\sum\limits_{n=0}^\infty \frac{t_4^{n+1}}{(n+1)!}C_{1,n}(t_1,t_2,t_5;x)\right] ~,\nonumber\\
J^{(4)}_{\hbox{\includegraphics[scale=0.09]{ggoggles.pdf}}}=&~J^{(2)}_{\hbox{\includegraphics[scale=0.09]{ggoggles.pdf}}}+\int \dnu(x)\frac{1}{4\pi}\frac{\Lambda^2}{\alpha_3+\alpha_4}B_0(t_1,t_2,t_5;x)~, \nonumber\\
J^{(1)}_{\hbox{\includegraphics[scale=0.12]{gbonneige.pdf}}}=&\int \dnu(x)~\hk_{3,4}(x,x)\sum\limits_{n=0}^\infty \frac{t_4^n}{n!}E_{0,n}(t_1,t_2,t_5;x)~,\nonumber\\
J^{(2)}_{\hbox{\includegraphics[scale=0.12]{gbonneige.pdf}}}=&~J^{(1)}_{\hbox{\includegraphics[scale=0.12]{gbonneige.pdf}}}+\int \dnu(x)\frac{1}{4\pi}\frac{\Lambda^2}{\alpha_3+\alpha_4}D_0(t_1,t_2,t_5;x) ~,\nonumber\\
J^{(3)}_{\hbox{\includegraphics[scale=0.12]{gbonneige.pdf}}}=&-\int \dnu(x)~\hk_{3,4}(x,x)\sum\limits_{n=0}^\infty \frac{t_4^n}{n!}D_{n+1}(t_1,t_2,t_5;x) ~,\nonumber\\
J^{(1)}_{\hbox{\includegraphics[scale=0.1]{gcylindre.pdf}}}=& \int \dnu(x)~\hk_{1,2}(x,x)\sum\limits_{n,m\ge 0}\frac{t_2^n}{n!}\frac{t_5^m}{m!}E_{n,m}(t_4+t_5,t_3,t_6;x)~,\nonumber\\
J^{(2)}_{\hbox{\includegraphics[scale=0.1]{gcylindre.pdf}}}=&~J^{(1)}_{\hbox{\includegraphics[scale=0.1]{gcylindre.pdf}}}+2\int \dnu(x)\frac{1}{4\pi}\frac{\Lambda^2}{\alpha_1+\alpha_2}D_0(t_4+t_5,t_3,t_6;x)+J^{1,2}_{4,5} ~,\nonumber\\
J^{(3)}_{\hbox{\includegraphics[scale=0.1]{gcylindre.pdf}}}=& \int \dnu(x)~\hk_{1,2}(x,x)\sum\limits_{n,m\ge 0}\frac{t_2^{n+1}}{(n+1)!}\frac{t_5^m}{m!}E_{n,m+1}(t_4+t_5,t_3,t_6;x)+J^{(3)}_{\hbox{\includegraphics[scale=0.12]{gbonneige.pdf}}}~,\nonumber\\
J^{(4)}_{\hbox{\includegraphics[scale=0.1]{gcylindre.pdf}}}=& ~J^{(3)}_{\hbox{\includegraphics[scale=0.1]{gcylindre.pdf}}} +\int \dnu(x)\frac{1}{4\pi}\frac{\Lambda^2}{\alpha_4+\alpha_5}D_0(t_1+t_2,t_3,t_6;x)~,
\label{apx:redef_J_i}
\end{align}
where all the terms in  $\mathcal{O}(A\Lambda^{2}\ln A\Lambda^2) $ are discarded. The fact that the $\alpha_i$ (and thus $t_i$) are dummy variables that can be renamed and are symmetrized, has been used to simplify the writings of  $J^{(2)}_{\hbox{\includegraphics[scale=0.1]{gcylindre.pdf}}}$ and $J^{(3)}_{\hbox{\includegraphics[scale=0.1]{gcylindre.pdf}}}$. Finally, the term $J^{1,2}_{4,5}$ in $J^{(2)}_{\hbox{\includegraphics[scale=0.1]{gcylindre.pdf}}}$ is the term proportional to $\Lambda^4$ defined in \eqref{eq:N2_integrals}. All contributions $\sim\Lambda^4$ have been discussed in section 3.2 and are summarized in \rtab{tab:N2_div}. At present we are only interested in the other types of divergences and thus we will simply drop the term $J^{1,2}_{4,5}$ in the following. 
We conjecture:
\begin{equation}
B_n(t_a,t_b,t_c;x)=
\left\{
\begin{array}{lc}
\frac{\left(-1\right)^n \Lambda^{2n}}{\left(4\pi\right)^2}(n-1)!\left[\frac{1}{\left(\alpha_a+\alpha_c\right)^{n}}+\frac{1}{\left(\alpha_b+\alpha_c\right)^{n}}\right]\ln A\Lambda^2+\mathcal{O}(\Lambda^{2n}) & \text{if}~ n\ge 1 ~,\\
\\
\frac{1}{(4\pi)^2}\left(\ln A\Lambda^2\right)^2+\mathcal{O}(\ln A\Lambda^2) & \text{if} ~n=0~.
\end{array}
\right.
\label{apx:bn}
\end{equation}
$\alpha_a$ and $\alpha_b$ being dummy variables, this may be rewritten as
\begin{equation}
B_n(t_a,t_b,t_c;x)=
\left\{
\begin{array}{lc}
2~\frac{\left(-1\right)^n (n-1)!}{\left(4\pi\right)^2}\left(\frac{\Lambda^{2}}{\alpha_b+\alpha_c}\right)^{n}\ln A\Lambda^2+\mathcal{O}(\Lambda^{2n}) & \text{if}~ n\ge 1 ~,\\
\\
\frac{1}{(4\pi)^2}\left(\ln A\Lambda^2\right)^2+\mathcal{O}(\ln A\Lambda^2) & \text{if} ~n=0~.
\end{array}
\right.
\label{apx:bn_sym}
\end{equation}
Likewise,
\begin{equation}
D_n(t_a,t_b,t_c;x)=
\left\{
\begin{array}{lc}
\frac{\left(-1\right)^n (n-1)!}{\left(4\pi\right)^2}\left(\frac{\Lambda^{2}}{\alpha_b+\alpha_c}\right)^{n}\ln A\Lambda^2+\mathcal{O}(\Lambda^{2n}) & \text{if}~ n\ge 1 ~,\\
\\
\frac{1}{(4\pi)^2}\left(\ln A\Lambda^2\right)^2+\mathcal{O}(\ln A\Lambda^2) & \text{if} ~n=0~.
\end{array}
\right.
\label{apx:dn}
\end{equation}
and
\begin{align}
C_{n,m}(t_a,t_b,t_c;x)&=~E_{n,m}(t_a,t_b,t_c;x)=~\frac{\left(-1\right)^{n+m}}{\left(4\pi\right)^2} (n+m)!\left(\frac{\Lambda^2}{\alpha_b+\alpha_c}\right)^{n+m+1}\ln A\Lambda^2+\mathcal{O}(\Lambda^{2n+2m+2})~.
\label{apx:cnm_enm}
\end{align}
From \eqref{eq:eqheatkernel}, one observes that
\begin{align}
C_{1,n}(t_a,t_b,t_c;x)=\frac{\mathrm{d}}{\mathrm{d}t_b}C_{0,n}(t_a,t_b,t_c;x)=-\frac{\mathrm{d}^2}{\mathrm{d}t_b^2}B_n(t_a,t_b,t_c;x)~,~~E_{0,n}(t_a,t_b,t_c;x)=-\frac{\mathrm{d}}{\mathrm{d}t_b}D_n(t_a,t_b,t_c;x)~,
\label{apx:rel_int_ref}
\end{align}
which is verified by the above expressions, before considering the symmetries between the $\alpha_i$. Putting everything together and remembering once more that the $\alpha_i$ are dummy variables, one gets:
\begin{align}
J^{(1)}_{\hbox{\includegraphics[scale=0.09]{ggoggles.pdf}}}&=\frac{A\Lambda^2}{\left(4\pi\right)^3}\left(\ln A\Lambda^2\right)^2\frac{1}{\alpha_1+\alpha_2+\alpha_3}~, \nonumber\\
J^{(1)}_{\hbox{\includegraphics[scale=0.12]{gbonneige.pdf}}}&=~J^{(3)}_{\hbox{\includegraphics[scale=0.12]{gbonneige.pdf}}}=~\frac{1}{2}~ J^{(2)}_{\hbox{\includegraphics[scale=0.09]{ggoggles.pdf}}}=~J^{(1)}_{\hbox{\includegraphics[scale=0.09]{ggoggles.pdf}}}~, \nonumber\\
J^{(3)}_{\hbox{\includegraphics[scale=0.09]{ggoggles.pdf}}}&=~J^{(2)}_{\hbox{\includegraphics[scale=0.12]{gbonneige.pdf}}}=~J^{(1)}_{\hbox{\includegraphics[scale=0.09]{ggoggles.pdf}}}+\frac{A\Lambda^2}{\left(4\pi\right)^3}\left(\ln A\Lambda^2\right)^2\frac{1}{\alpha_1+\alpha_2}~, \nonumber\\
J^{(4)}_{\hbox{\includegraphics[scale=0.09]{ggoggles.pdf}}}&=~2J^{(1)}_{\hbox{\includegraphics[scale=0.09]{ggoggles.pdf}}}+\frac{A\Lambda^2}{\left(4\pi\right)^3}\left(\ln A\Lambda^2\right)^2\frac{1}{\alpha_1+\alpha_2}~, \nonumber\\
J^{(3)}_{\hbox{\includegraphics[scale=0.1]{gcylindre.pdf}}}&=~J^{(1)}_{\hbox{\includegraphics[scale=0.1]{gcylindre.pdf}}}=~\frac{A\Lambda^2}{\left(4\pi\right)^3}\left(\ln A\Lambda^2\right)^2\frac{1}{\alpha_1+\alpha_2+\alpha_3+\alpha_4}~,\nonumber\\
J^{(2)}_{\hbox{\includegraphics[scale=0.1]{gcylindre.pdf}}}&=~J^{(1)}_{\hbox{\includegraphics[scale=0.1]{gcylindre.pdf}}}+2\frac{A\Lambda^2}{\left(4\pi\right)^3}\left(\ln A\Lambda^2\right)^2\frac{1}{\alpha_1+\alpha_2}~\nonumber\\
J^{(4)}_{\hbox{\includegraphics[scale=0.1]{gcylindre.pdf}}}&=~J^{(1)}_{\hbox{\includegraphics[scale=0.1]{gcylindre.pdf}}}+\frac{A\Lambda^2}{\left(4\pi\right)^3}\left(\ln A\Lambda^2\right)^2\frac{1}{\alpha_1+\alpha_2}~,
\label{apx:int_ref}
\end{align}
up to subleading divergences. \\

One encounters also integrals such as 
\begin{align}
L_{\hbox{\includegraphics[scale=0.12]{gbonneige.pdf}}}=&\int \dnu(x)\dnu(y)\dnu(z)~\hk_1(z,z)\hk_2(x,y)\hk_3(x,y)\kt_4(x,z)\left(-\frac{\mathrm{d}}{\mathrm{d}t_5}\kt_5(y,z)\right)
\label{apx:L_neige}
\end{align}
whose explicit computation requires to Taylor expand a product of two $\hk$ or $\kt$. We denote such integrals by $L$.
Integrating over $z$ around $x$ through the exponential term in $\kt_4(x,z)$, see \eqref{eq:k_hk_xy}, leads us to Taylor expand $\hk_1(z,z)\frac{\mathrm{d}}{\mathrm{d}t_5}\kt_5(y,z)$ in $(z-x)$ around $x$. 
After integration, one gets terms such as:
\begin{align}
\int \dnu(x)\dnu(y)~\hk_2(x,y)\hk_3(x,y)\partial^x_{i_1}\dots\partial^x_{i_r}\hk_1(x,x)\partial^x_{j_1}\dots\partial^x_{j_s}\left(-\frac{\mathrm{d}}{\mathrm{d}t_5}\kt_5(x,y)\right)
\label{apx:drkxx_dskxy}
\end{align}
with $r+s$ even. If $s$ is odd, then the function $\hk_2(x,y)\hk_3(x,y)\partial^x_{j_1}\dots\partial^x_{j_s}\left(-\frac{\mathrm{d}}{\mathrm{d}t_5}\kt_5(x,y)\right)$ is odd and performing the integral over $y$ kills the contribution: $r$ and $s$ have to be even. Since $\kt(t,x,x)$ depends on $x$ only through $\tilde{G}_\zeta^{A_0}(x)$, see \eqref{eq:hk_xx}, for $r\ge 2$, $\partial^x_{i_1}\dots\partial^x_{i_r}\hk_1(x,x)$ does not contribute to the leading divergence by neither a factor $\ln A\Lambda^2$ nor $\Lambda^2$. The diverging contributions may thus only come from the integral over $y$. However, applying $s$ derivatives on $\frac{\mathrm{d}}{\mathrm{d}t_5}\kt_5(x,y)$ for any even $s$ leads to terms similar to $\left(-1\right)^{\frac{s}{2}}\left(\frac{\mathrm{d}}{\mathrm{d}t_5}\right)^{1+\frac{s}{2}}\kt_5(x,y)$. Integrating over $y$ one gets $B_{1+\frac{s}{2}}(t_2,t_3,t_5;x)$ which only produces one of the two $\ln A\Lambda^2$ of the leading divergence. The only terms contributing to $A\Lambda^2\left(\ln A\Lambda^2\right)^2$ are thus the terms with $r=0$ and $s$ even. 

Thus, up to subleading divergences, the previous integral gives:
\begin{align}
L_{\hbox{\includegraphics[scale=0.12]{gbonneige.pdf}}}=&-\int \dnu(x)~\hk_{1}(x,x)\sum\limits_{n=0}^\infty \frac{t_4^{n}}{n!}B_{n+1}(t_2,t_3,t_5;x)=~J^{(2)}_{\hbox{\includegraphics[scale=0.09]{ggoggles.pdf}}}~.
\label{apx:L_neige_serie}
\end{align}
Similarly, one may compute
\begin{align}
L^{(1)}_{\hbox{\includegraphics[scale=0.1]{gcylindre.pdf}}}=&\int \dnu(x)\dnu(y)\dnu(z)\dnu(w)~\hk_{1}(x,y)\hk_2(x,y)\kt_{3}(y,z)\hk_{4}(z,w)\kt_5(z,w)\left(-\frac{\mathrm{d}}{\mathrm{d}t_6}\kt_6(x,w)\right)~, \nonumber\\
=&-\int \dnu(x)\dnu(y)\dnu(z)~\hk_{1}(x,y)\hk_2(x,y)\kt_{3}(y,z)\hk_{4,5}(z,z)\sum\limits_{n=0}^\infty \frac{t_5^{n}}{n!}\left(\frac{\mathrm{d}}{\mathrm{d}t_6}\right)^{n+1}\kt_6(x,z)~, \nonumber\\
=&-\int \dnu(x)\dnu(y)~\hk_{1}(x,y)\hk_2(x,y)\hk_{4,5}(y,y)\sum\limits_{n,m\ge0} \frac{t_3^{m}}{m!}\frac{t_5^{n}}{n!}\left(\frac{\mathrm{d}}{\mathrm{d}t_6}\right)^{n+m+1}\kt_6(x,y)~, \nonumber\\
=&~~2~J^{(1)}_{\hbox{\includegraphics[scale=0.1]{gcylindre.pdf}}}, \nonumber\\
L^{(2)}_{\hbox{\includegraphics[scale=0.1]{gcylindre.pdf}}}=&\int \dnu(x)\dnu(y)\dnu(z)\dnu(w)~\hk_{1}(x,y)\hk_2(x,y)\kt_{3}(y,z)\hk_{4}(z,w)\left(-\frac{\mathrm{d}}{\mathrm{d}t_5}\kt_5(z,w)\right)\kt_6(x,w)~, \nonumber\\
=&~L^{(1)}_{\hbox{\includegraphics[scale=0.1]{gcylindre.pdf}}}+\frac{1}{4\pi}\frac{\Lambda^2}{\alpha_4+\alpha_5}~\sum\limits_{n=0}^\infty \frac{t_6^{n}}{n!}\int \dnu(x)B_{n}(t_1,t_2,t_3;x)~,  \nonumber\\
=&~L^{(1)}_{\hbox{\includegraphics[scale=0.1]{gcylindre.pdf}}}+\frac{A\Lambda^2}{\left(4\pi\right)^3}\left(\ln A\Lambda^2\right)^2\frac{1}{\alpha_4+\alpha_5}~,
\label{apx:L_int}
\end{align}
up to $\mathcal{O}(A\Lambda^{2}\ln A\Lambda^2) $ terms.

Since at least two $\hk$s are needed to obtain the two $\ln A\Lambda^2$ of the leading divergence, it is easy to compute
\begin{align}
L^{(1)}_{\hbox{\includegraphics[scale=0.1]{gtetra.pdf}}}&= \int \dnu(x)\dnu(y)\dnu(z)\dnu(w)~\hk_{1}(x,w)\hk_2(y,w)\kt_{3}(z,w)\kt_{4}(x,y)\kt_5(x,z)\kt_6(y,z)~.
\label{apx:L1_tetra_def}
\end{align}
Indeed, integrating over $w$ through $\kt_3(z,w)$ requires to Taylor expand $\hk_{1}(x,w)\hk_2(y,w)$. The only term in this expansion keeping the structure of the $\hk$s and thus contributing to the leading divergence is the first one: $\hk_{1}(x,z)\hk_2(y,z)$. Thus,
\begin{align}
L^{(1)}_{\hbox{\includegraphics[scale=0.1]{gtetra.pdf}}}&= \int \dnu(x)\dnu(y)\dnu(z)~\hk_{1}(x,z)\hk_2(y,z)\kt_{4}(x,y)\kt_5(x,z)\kt_6(y,z)+\mathcal{O}(A\Lambda^{2}\ln A\Lambda^2) \nonumber\\
&=~J^{(1)}_{\hbox{\includegraphics[scale=0.09]{ggoggles.pdf}}}+\mathcal{O}(A\Lambda^{2}\ln A\Lambda^2)~.
\label{apx:L1_tetra}
\end{align}
Similarly, in
\begin{align}
L^{(2)}_{\hbox{\includegraphics[scale=0.1]{gtetra.pdf}}}&= \int \dnu(x)\dnu(y)\dnu(z)\dnu(w)~\hk_{1}(x,w)\kt_2(y,w)\kt_{3}(z,w)\kt_{4}(x,y)\kt_5(x,z)\hk_6(y,z)~,
\label{apx:L2_tetra_def}
\end{align}
if any partial derivative acts on one of the two $\hk$s through the Taylor expansion, the $\left(\ln A\Lambda^2\right)^2$ are lost. Thus,
\begin{align}
L^{(2)}_{\hbox{\includegraphics[scale=0.1]{gtetra.pdf}}}&= \int \dnu(x)\dnu(y)\dnu(z)~\kt_{4}(x,y)\kt_5(x,z)\hk_6(y,z)\hk_{1}(x,z)\sum\limits_{n=0}^\infty\frac{t_3^n}{n!}\frac{\mathrm{d}^n}{\mathrm{d}t_2^n}\kt_2(y,z)~, \nonumber\\
&= \int \dnu(x)~\hk_{1,5}(x,x)\sum\limits_{n,m\ge 0}\frac{t_3^n}{n!}\frac{t_5^m}{m!}C_{n,m}(t_6,t_2,t_4;x)=~J^{(1)}_{\hbox{\includegraphics[scale=0.1]{gcylindre.pdf}}}~,
\label{apx:L2_tetra}
\end{align} 
up to subleading terms. It is possible to compute
\begin{align}
L^{(3)}_{\hbox{\includegraphics[scale=0.1]{gtetra.pdf}}}&= \int \dnu(x)\dnu(y)\dnu(z)\dnu(w)~\hk_1(x,y)\kt_2(x,z)\left(-\frac{\mathrm{d}}{\mathrm{d}t_3}\kt_{3}(y,z)\right)\hk_{4}(x,w)\hk_5(y,w)\kt_{6}(z,w)~, \nonumber\\
L^{(4)}_{\hbox{\includegraphics[scale=0.1]{gtetra.pdf}}}&= \int \dnu(x)\dnu(y)\dnu(z)\dnu(w)~\kt_{1}(x,y)\hk_2(x,z)\left(-\frac{\mathrm{d}}{\mathrm{d}t_3}\kt_{3}(y,z)\right)\hk_{4}(x,w)\hk_5(y,w)\kt_6(z,w)~, \nonumber\\
L^{(5)}_{\hbox{\includegraphics[scale=0.1]{gtetra.pdf}}}&= \int \dnu(x)\dnu(y)\dnu(z)\dnu(w)~\left(-\frac{\mathrm{d}}{\mathrm{d}t_3}\kt_{1}(x,y)\right)\hk_2(x,z)\kt_3(y,z)\hk_{4}(x,w)\hk_5(y,w)\kt_{6}(z,w)~,
\label{apx:L345_tetra_def}
\end{align}
with the same reasoning. In these three integrals, integrating over $w$ implies to Taylor expand a product of two $\hk$s. Since there are only three $\hk$s in these integrals, this means that, if one wants to extract the leading divergence with two $\ln A\Lambda^2$, one shall only keep in the Taylor expansion the terms with the derivatives acting on one $\hk$. Up to subleading terms, one has
\begin{align}
L^{(3)}_{\hbox{\includegraphics[scale=0.1]{gtetra.pdf}}}&= ~J^{(3)}_{\hbox{\includegraphics[scale=0.09]{ggoggles.pdf}}}+L^{(3a)}_{\hbox{\includegraphics[scale=0.1]{gtetra.pdf}}}+L^{(3b)}_{\hbox{\includegraphics[scale=0.1]{gtetra.pdf}}}~, \nonumber\\
L^{(3a)}_{\hbox{\includegraphics[scale=0.1]{gtetra.pdf}}}&=\int \dnu(x)\dnu(y)\dnu(z)~\hk_1(x,y)\kt_2(x,z)\frac{\mathrm{d}}{\mathrm{d}t_3}\kt_{3}(y,z)\hk_{4}(x,z)\sum\limits_{n=0}^\infty\frac{t_6^{n+1}}{(n+1)!}\frac{\mathrm{d}^n}{\mathrm{d}t_5^n}\kt_5(y,z)~, \nonumber\\
L^{(3b)}_{\hbox{\includegraphics[scale=0.1]{gtetra.pdf}}}&=\int \dnu(x)\dnu(y)\dnu(z)~\hk_1(x,y)\kt_2(x,z)\frac{\mathrm{d}}{\mathrm{d}t_3}\kt_{3}(y,z)\hk_5(y,z)\sum\limits_{n=0}^\infty\frac{t_6^{n+1}}{(n+1)!}\frac{\mathrm{d}^n}{\mathrm{d}t_4^n}\kt_{4}(x,z)~.
\label{apx:L3_tetra}
\end{align}
Integrating over $x$ in $L^{(3a)}_{\hbox{\includegraphics[scale=0.1]{gtetra.pdf}}}$ leads to 
\begin{align}
&\int \dnu(z)~\hk_{2+4}(z,z)\sum\limits_{n=0}^\infty\frac{t_6^{n+1}}{(n+1)!}C_{1,n}(t_1,t_3,t_5;z)=~J^{(1)}_{\hbox{\includegraphics[scale=0.09]{ggoggles.pdf}}}-\frac{A\Lambda^2}{\left(4\pi\right)^3}\left(\ln A\Lambda^2\right)^2\frac{1}{\alpha_1+\alpha_2}~,
\end{align}
while integrating over $y$ in $L^{(3b)}_{\hbox{\includegraphics[scale=0.1]{gtetra.pdf}}}$ transforms one of the two remaining $\hk$s, resulting in a $\mathcal{O}(A\Lambda^{2}\ln A\Lambda^2)$ term. $L^{(4)}_{\hbox{\includegraphics[scale=0.1]{gtetra.pdf}}}$ and $L^{(5)}_{\hbox{\includegraphics[scale=0.1]{gtetra.pdf}}}$ are computed in a similar way. Summing up: 
\begin{align}
L^{(3)}_{\hbox{\includegraphics[scale=0.1]{gtetra.pdf}}}= ~2J^{(1)}_{\hbox{\includegraphics[scale=0.09]{ggoggles.pdf}}}~, ~~~~~~L^{(4)}_{\hbox{\includegraphics[scale=0.1]{gtetra.pdf}}}=~J^{(1)}_{\hbox{\includegraphics[scale=0.1]{gcylindre.pdf}}}+2J^{(1)}_{\hbox{\includegraphics[scale=0.09]{ggoggles.pdf}}}~,~~~~~~ L^{(5)}_{\hbox{\includegraphics[scale=0.1]{gtetra.pdf}}}=~J^{(1)}_{\hbox{\includegraphics[scale=0.1]{gcylindre.pdf}}}+J^{(1)}_{\hbox{\includegraphics[scale=0.09]{ggoggles.pdf}}}~.
\label{apx:L345_tetra}
\end{align}

The same idea is used to compute 
\begin{align}
L^{(3)}_{\hbox{\includegraphics[scale=0.1]{gcylindre.pdf}}}&=\int \dnu(x)\dnu(y)\dnu(z)\dnu(w)~\hk_1(x,z)\hk_2(x,z)\left(-\frac{\mathrm{d}}{\mathrm{d}t_3}\kt_3(y,z)\right)\hk_4(y,w)\hk_5(y,w)\left(-\frac{\mathrm{d}}{\mathrm{d}t_6}\kt_6(x,w)\right)~, \nonumber\\
L^{(6)}_{\hbox{\includegraphics[scale=0.1]{gtetra.pdf}}}&=\int \dnu(x)\dnu(y)\dnu(z)\dnu(w) ~\hk_1(x,y)\hk_2(x,z)\left(-\frac{\mathrm{d}}{\mathrm{d}t_3}\kt_3(y,z)\right)\hk_4(z,w)\hk_5(y,w)\left(-\frac{\mathrm{d}}{\mathrm{d}t_6}\kt_6(x,w)\right)~,
\label{apx:S5_P9}
\end{align}
the requirement is then to leave two of the four $\hk$s without derivatives. Among such terms, one gets also subleading terms not considered here. In the end, one obtains:
\begin{align}
L^{(3)}_{\hbox{\includegraphics[scale=0.1]{gcylindre.pdf}}}&=~2L^{(6)}_{\hbox{\includegraphics[scale=0.1]{gtetra.pdf}}}=~4J^{(1)}_{\hbox{\includegraphics[scale=0.1]{gcylindre.pdf}}}~.
\label{apx:S5_P9_res}
\end{align}

\end{appendix}

\newpage

\bibliographystyle{unsrt}

\end{document}